\documentclass{article}

\usepackage{arxiv}
\usepackage[utf8]{inputenc} 
\usepackage[T1]{fontenc}    
\usepackage{hyperref}       
\usepackage{url}            
\usepackage{booktabs}       
\usepackage{amsfonts}       
\usepackage{nicefrac}       
\usepackage{microtype}      
\usepackage{lipsum}
\usepackage{graphicx}
\usepackage{subcaption}
\usepackage{amsmath,amsfonts,amssymb}
\usepackage{comment}
\graphicspath{{Figures/}}
\usepackage{color}
\usepackage[algo2e,ruled,vlined,linesnumbered]{algorithm2e}
\usepackage{mdframed}
\usepackage{comment}
\usepackage{multirow}
\usepackage{bbm}

\newcommand{\mbf}{\mathbf}
\newcommand{\mbb}{\mathbb}
\newcommand{\mcl}{\mathcal}
\newcommand{\f}{\frac}
\newcommand{\bs}{\boldsymbol}
\newcommand{\RNum}[1]{\uppercase\expandafter{\romannumeral #1\relax}}
\newcommand{\R}{\mathbb{R}}

\newcommand{\p}{\partial}
\newcommand{\T}{\textnormal}
\newcommand{\x}{\mathbf{x}}

\newtheorem{theorem}{Theorem}[section]
\newtheorem{remark}[theorem]{Remark}

\title{Enhanced data efficiency using deep neural networks and Gaussian processes for aerodynamic design optimization}

\author{
  S. Ashwin Renganathan \\
  Mathematics and Computer Science Division\\
  Argonne National Laboratory\\
  Lemont, IL 60439, USA\\
  \texttt{srenganathan@anl.gov} \\
    \And
   Romit Maulik \\
  Argonne Leadership Computing Facility\\
  Argonne National Laboratory\\
  Lemont, IL 60439, USA\\   \texttt{rmaulik@anl.gov} \\
   \And
   Jai Ahuja \\
   School of Aerospace Engineering \\
   Georgia Institute of Technology\\
   Atlanta, GA 30313, USA \\
   \texttt{jai.ahuja@gatech.edu} \\
}

\begin{document}
\maketitle

\begin{abstract}
Adjoint-based optimization methods are attractive for aerodynamic shape design primarily due to their computational costs being independent of the dimensionality of the input space and their ability to generate high-fidelity gradients that can then be used in a gradient-based optimizer. This makes them very well suited for high-fidelity simulation based aerodynamic shape optimization of highly parametrized geometries such as aircraft wings. However, the development of adjoint-based solvers involve careful mathematical treatment and their implementation require detailed software development. Furthermore, they can become prohibitively expensive when multiple optimization problems are being solved, each requiring multiple restarts to circumvent local optima. In this work, we propose a machine learning enabled, surrogate-based framework that replaces the expensive adjoint solver, without compromising on predicting predictive accuracy. Specifically, we first train a deep neural network (DNN) from training data generated from evaluating the high-fidelity simulation model on a model-agnostic, design of experiments on the geometry shape parameters. The optimum shape may then be computed by using a gradient-based optimizer coupled with the trained DNN. Subsequently, we also perform a gradient-free Bayesian optimization, where the trained DNN is used as the prior mean. We observe that the latter framework (DNN-BO) improves upon the DNN-only based optimization strategy for the same computational cost. Overall, this framework predicts the true optimum with very high accuracy, while requiring far fewer high-fidelity function calls compared to the adjoint-based method. Furthermore, we show that multiple optimization problems can be solved with the same machine learning model with high accuracy, to amortize the offline costs associated with constructing our models. Our methodology finds applications in the early stages of aerospace design.
\end{abstract}


\section{Introduction}
\label{s:Introduction}
Aerospace design optimization, specifically aerodynamic shape optimization (ASO) and multidisciplinary design optimization (MDO), involves the optimization of expensive-to-evaluate functions in a high dimensional decision space. Additionally, the functions and associated constraints can potentially be nonlinear, non-convex and, available as a (black-box) simulation, making optimization challenging. These simulations (a.k.a high-fidelity models) are computationally expensive because they require iterative numerical solutions to a system of discretized coupled nonlinear partial differential equations (PDEs) with very large degrees of freedom; e.g., computational fluid dynamics (CFD) and computational structural dynamics (CSD) models for practical engineering problems typically involve $\mcl{O}(10^6)$ degrees of freedom.
This make them prohibitively expensive when used in an optimization setting. We present data-efficient strategies by utilizing state-of-the-art machine learning (ML) to cope with both the cost of expensive simulations as well as the difficulty in optimizing quantities of interest (QoIs) derived from the simulations.

We specifically address the ASO problem, where the optimal aerodynamic shape (of, e.g., aircraft wing or wing section) is sought, such that it achieves the best overall aerodynamic performance such as minimum drag, given certain design constraints and a simulation that maps the design variables to the aerodynamic performance. Such a problem is efficiently solved via the adjoint method~\cite{jameson1988aerodynamic, jameson1998optimum, jameson1989computational} which treats the parameters of the aerodynamic shape as control parameters and computes the gradient of a suitably chosen cost function (e.g. drag or the lift-to-drag ratio) in terms of the control parameters which can then be used within a gradient-based optimizer. The adjoint-based method is attractive due to its computational cost being independent of the design space dimension and has been successfully applied toward transonic airfoil design~\cite{jameson1989computational}, transonic wing design~\cite{jameson1994optimum} and entire aircraft configurations such as the NASA common research model (CRM)~\cite{chen2016aerodynamic} and the blended wing-body~\cite{lyu2014aerodynamic}. However, adjoint-based ASO poses the following challenges. Firstly, the approach requires the derivatives of the cost function in terms of the flow-field variables; this essentially makes it \emph{intrusive} in the sense that simulation codes have to be structurally modified by differentiating almost every line of code to achieve this. Secondly, the computation of the adjoint-based derivatives involve the numerical solution of the adjoint equations which are another set of PDEs whose specification of boundary conditions and numerical solution require careful mathematical treatment \cite{jameson1988aerodynamic} and they are known to be often difficult to converge displaying high sensitivity to the computational mesh quality. Thirdly, the cost of the adjoint-based approach can become prohibitive when multiple optimization problems are required to be solved. For example, in the design of an aircraft wing, consider the following optimization problem objectives: (i) \emph{P1}: minimize drag with wing-thickness constraints, (ii) \emph{P2}: minimize drag with wing-thickness and lift constraint, (iii) \emph{P3}: minimize drag with wing-thickness and trim constraints and (iv) \emph{P4}: maximize lift-to-drag ratio with wing-thickness constraints. Solving \emph{P1} through \emph{P4} with an adjoint-based method requires independent solutions which can collectively cost in $\mcl{O}(100)$ high-fidelity function evaluations that can prove to be prohibitive when each evaluation is computationally expensive. Additionally, across the aforementioned optimization problems, there is potential for information reuse, which is not realized when solving each optimization problem independently. This is because, the optimization search paths across \emph{P1-P4} can overlap and hence the expensive evaluations can be reused. In this work, we seek purely data-driven and data-efficient strategies to solve aerodynamic design optimization problems that can potentially circumvent the aforementioned challenges and yet result in high quality designs.

To address the limitations of the intrusive and expensive yet accurate adjoint-based method for the ASO problem, we propose a deep neural network (DNN) based approach where a DNN is trained to emulate the aerodynamic quantities of interest (e.g. lift and drag coefficients) as a function of the design parameters which in this work are parameters that control the aerodynamic shape. Then we show that a gradient-based optimization with the DNN emulator, without any further expensive function queries, can produce accurate results for the ASO problem. Additionally, we propose a DNN-enhanced Gaussian process (GP) model (DNN-GP) where a DNN learned is used as the prior mean of the GP to generate a probabilistic emulator of the high-fidelity model whose aposteriori mean interpolates the training dataset. By using the trained DNN-GP in a Bayesian optimization framework, we show that the ASO problem can be solved more accurately than using the DNN alone, with only a few dozen high-fidelity function evaluations. 

The idea of employing a cheap-to-evaluate \emph{surrogate} model to bring down overall computational costs in simulation optimization has been applied widely in aerospace design, such as in aircraft engine nacelle shape optimization~\cite{song2007surrogate}, transonic wing design~\cite{han2015surrogate} and hypersonic wing design~\cite{liu2019surrogate} just to name a few. In all of these works, a response surface is built for the cost function which is then optimized in lieu of the actual cost function. The work in  \cite{koziel2013surrogate} uses a surrogate model to correct a biased lower fidelity model to improve the cost efficiency (in terms of the number of high-fidelity function evaluations) of the ASO. More recently, deep learning has been finding application in aerospace design optimization.\cite{yan2019aerodynamic} develop a multifidelity framework where deep belief network (DBN) is trained and used as a low-fidelity model and \cite{li2020efficient} use generative adversarial networks (GAN) to develop generative models for airfoil and wing shapes which are then used within an optimization framework. \cite{tao2019application} use a DBN to generate a linear regression surrogate in a multifidelity framework. Subsequently they use their linear model in a gradient-free particle swarm optimization framework. Neural networks have also been used directly in genetic and population-based optimization algorithms quite frequently \cite{hacioglu2007fast,shahrokhi2010surrogate,ribeiro2012airfoil,kotinis2012multi,pehlivanoglu2019efficient}. In recent times, the power of backpropagation has led to the efficient integration of neural network surrogates within gradient-based optimizers. \cite{bouhlel2020scalable} utilize deep fully connected artificial neural networks with training enhanced by adjoint information to obtain fast optimizers for airfoil shapes. \cite{hui2020fast}., use a deep convolutional neural network to make fast predictions of pressure coefficients for a variety of airfoils. \cite{sekar2019inverse} construct an inverse design process using convolutional neural networks where the input to a data-driven framework is given by a pressure coefficient distribution and the output is a generated airfoil shape  Similar frameworks have also been proposed in \cite{chiu2020airfoil,du2020b} where a generative process has leveraged GANs to construct novel airfoils constrained by some choice of a spline, given some reward. However, we demonstrate the data-efficiency of our approach with a relatively small number ($\mcl{O}(10)$) of high-fidelity evaluations generated apriori. We show that we can train accurate DNN models and further enhance their accuracy by hybridizing them with a GP model. Furthermore, we show that the trained models can be used to solve multiple optimization problems accurately within the domain of relevance of the surrogate models. We make direct comparisons to the true optimized shape and the adjoint method to validate our claims.


Our main contributions are as follows. We show that with a carefully trained DNN model, the mapping between the aerodynamic shape parameters and the aerodynamic quantities of interest (e.g. lift and drag coefficients) can be learned quite accurately such that the trained DNN can be directly used within a multistart gradient-based local optimizer to find the optimal design. This way, we demonstrate a data-efficient yet accurate strategy for aerodynamic design optimization. Secondly, we further improve the accuracy of the approach by combining the DNN with a GP model. More specifically, whereas the DNN learns the overall trends in the data, the GP learns the local volatility that the DNN fails to capture. This way, a hybrid, probabilistic DNN-GP model is constructed whose expectation interpolates the data and which is then used in a Bayesian optimization framework to find the globally optimum shape within a constrained space. The resulting DNN-BO framework is demonstrated to be more data-efficient and accurate than the DNN alone. Finally, we demonstrate a framework (DNN-BOc) that uses the same trained models on a constrained optimization problem where the constraint evaluations are also expensive. Finally, we make direct comparisons against the adjoint-based optimization to demonstrate the efficiency of our methods.

We make the following remarks about the overarching goals of this paper.
\begin{remark}
    The adjoint-based method for ASO uses high-fidelity evaluations of the objective and gradients and are more theoretically grounded and well established. However, development of an adjoint-based framework involves detailed software development. On the other hand, machine learning models and gradient-based optimizers are both easy to implement and are freely available. Therefore this work aims to investigate the feasibility of a data-driven approach when sophisticated adjoint-based solvers are unavailable. 
\end{remark}

\begin{remark}
    The adjoint-based method for ASO is highly efficient, particularly, as the input dimension size is increases. On the other hand, it becomes increasingly prohibitive to fit machine learning models as the input dimension size increases. Whereas this work is focused only on input dimension size in $\mcl{O}(10)$, we identify some potential directions to handle larger sized problems in section~\ref{s:conclusion}. 
\end{remark}

The rest of the article is organized as follows. The following section presents the main problem formulation including the high-fidelity model, the aerodynamic shape parametrization
and the optimization problem statements. Following that, in section~\ref{s:opt_strats} we discuss the various strategies for ASO that are studied in this work including the proposed DNN, DNN-BO and DNN-BOc frameworks. We present the results in section~\ref{s:results} following which the conclude the paper with a brief summary and an outlook on future work.

\section{Problem formulation}
\label{s:problem_formulation}
Our methodology is demonstrated on the ASO problem of identifying the minimum drag shape of the RAE2822 airfoil under transonic flight conditions. We begin by presenting the compressible Euler equations governing the inviscid transonic flow as well the parametrization used to describe the airfoil shape. Then we discuss the actual optimization problems we seek to solve with our proposed approach.

\subsection{Compressible Euler equations}
The Euler equation governing the two-dimensional, compressible inviscid flow, in conservation form is given as
\begin{equation}
\nabla_x \mbf{F} + \nabla_y \mbf{G} = 0, 
\label{e:Euler_cons}
\end{equation}

where
\begin{equation*}
\begin{aligned}
&& \mbf{F} = 
\begin{bmatrix}
\rho u \\
\rho u^2 + p \\
\rho uv \\
\rho uH
\end{bmatrix},~ \mbf{G} = \begin{bmatrix}
\rho v \\
\rho uv \\
\rho v^2 + p \\
\rho vH
\end{bmatrix} \\
&& H = E + \frac{p}{\rho}\\
&& \rho E = \frac{1}{2} \rho (u^2 + v^2) + \frac{p}{\gamma -1}.
\end{aligned}
\end{equation*}
In \eqref{e:Euler_cons},  $\rho,u,v,$ and $p$ are the primitive variables, namely, the density, velocity components, and static pressure, respectively; $H$ is the enthalpy; $E$ is the internal energy and $\gamma$ is the ratio of specific heats. $\nabla_x$ and $\nabla_y$ are the $x$ and $y$ components of the gradient operator $\nabla$, respectively. The freestream ($\infty$) boundary conditions are specified with a flow direction ($\alpha$), the Mach number ($M$), $p$, and static temperature ($T$). The freestream boundary values on the boundary face of the computational domain are computed based on extrapolation of Riemann invariants under the assumption of irrotational, quasi-1D flow in the boundary-normal direction. The airfoil surface is modeled as an adiabatic slip wall where all the primitive and thermodynamic variables are extrapolated from the interior domain via reconstruction gradients. The numerical solution to the nonlinear system is obtained with a coupled implicit finite-volume based solver with second-order spatial discretization. The gradients are computed with the hybrid Gauss--least-squares method and the Venkatakrishnan limiter \cite{venkatakrishnan1995convergence}. 

\begin{figure}
  \begin{subfigure}{0.5\textwidth}
  \includegraphics[width=1\linewidth]{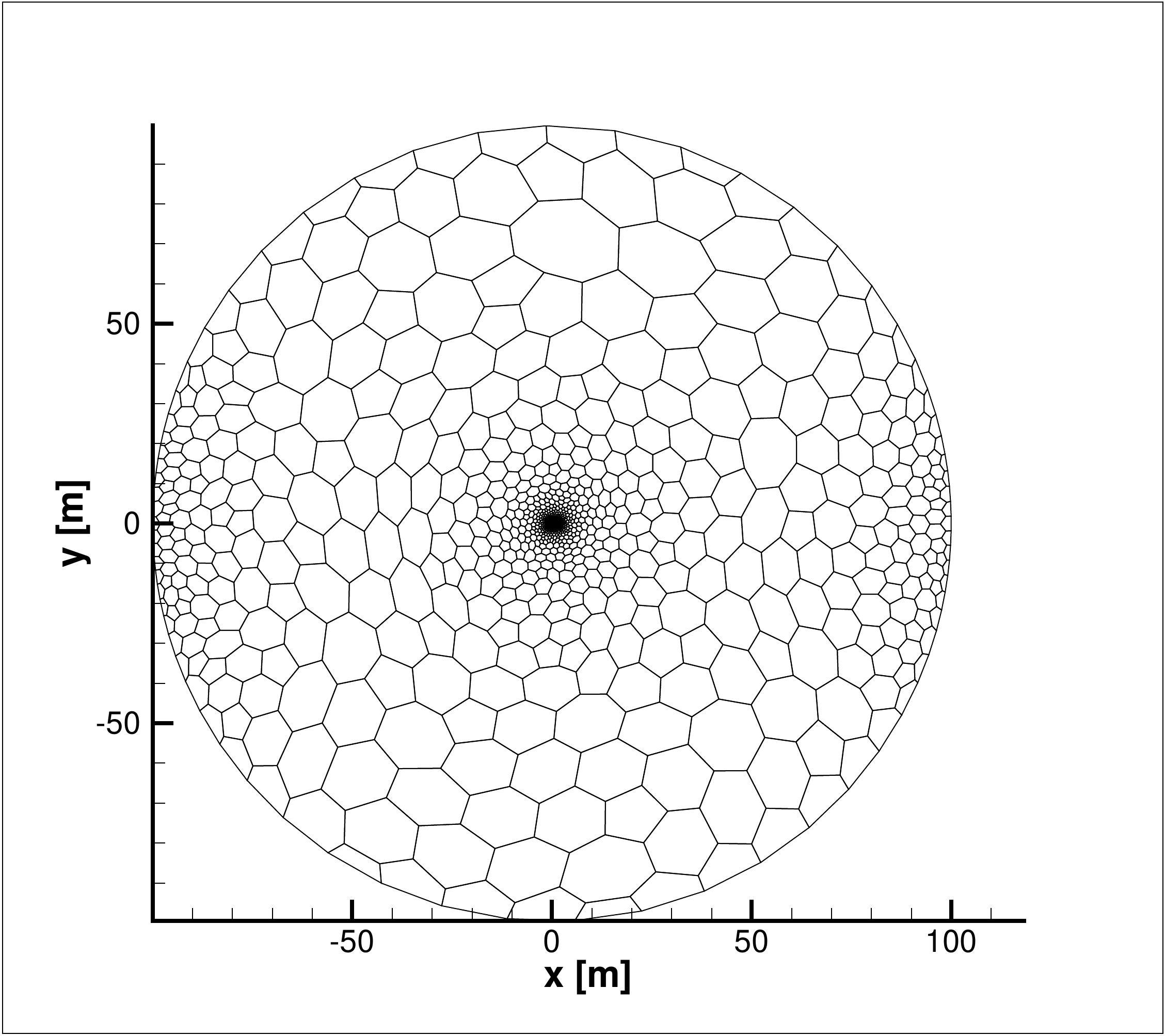}
  \caption{Flow domain with mesh}
  \label{f:RAE2822_Grid}
  \end{subfigure}%
  \begin{subfigure}{0.5\textwidth}
  \includegraphics[width=1\linewidth]{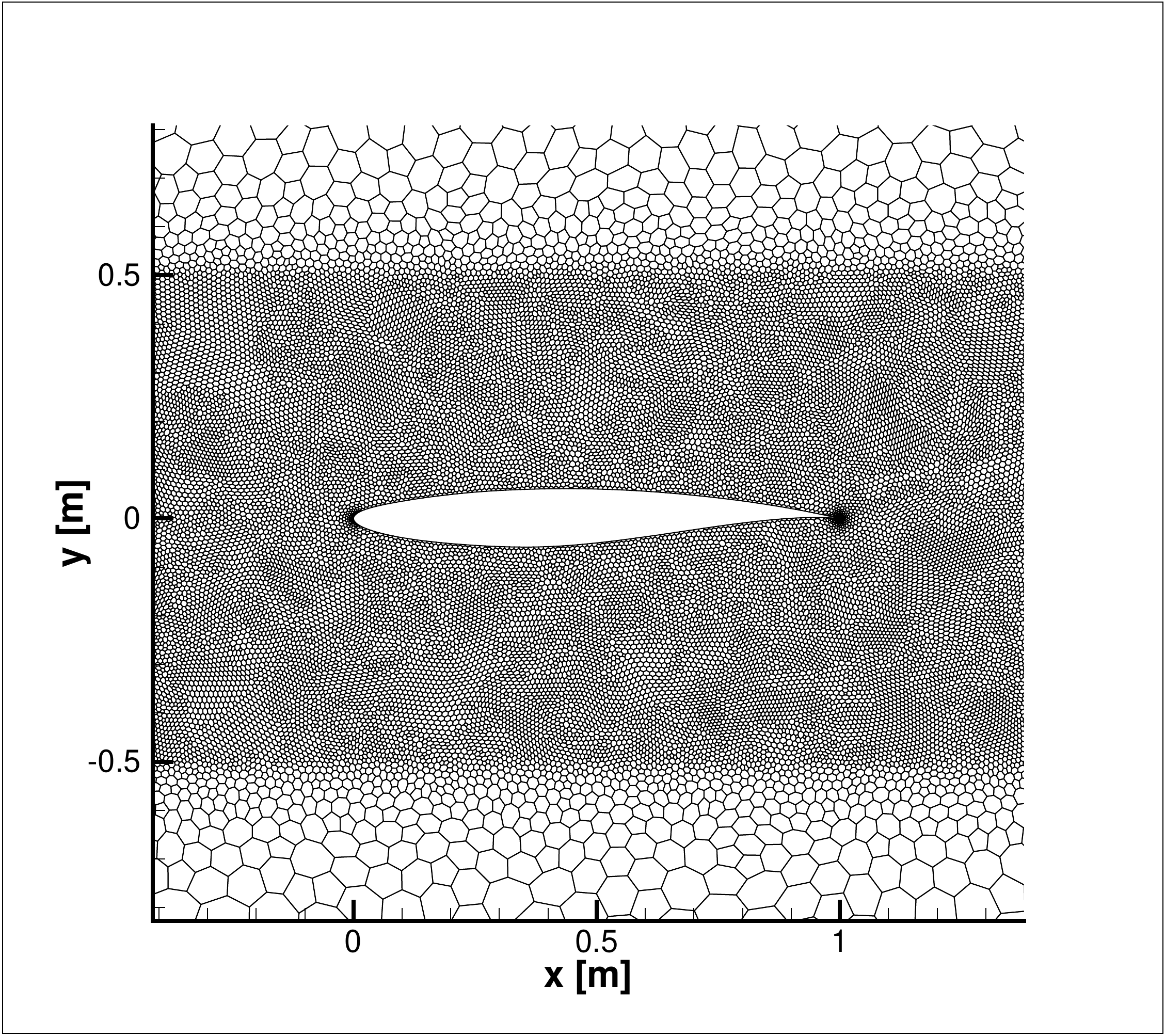}
  \caption{Near field mesh}
  \label{f:RAEMesh}
  \end{subfigure}%
  \caption{Farfield (a) and near field (b) computational grid for the airfoil test case}
  \label{f:airfoil_grid}
\end{figure}

The RAE2822 is a commonly used canonical test case for aerospace engineering problems under transonic flight conditions. A spherical domain of $100$ airfoil chord length radius is used to model the fluid domain, which is meshed with 27,857 polyhedral mesh elements with near-field refinement to capture the shock as shown in Figure~\ref{f:airfoil_grid}. In this work, the main quantities of interest are the lift, drag and pressure coefficients defined as
\[C_l = \frac{L}{\frac{1}{2}\rho_\infty (M_\infty\times a_\infty)^2} \quad 
C_d = \frac{D}{\frac{1}{2}\rho_\infty (M_\infty\times a_\infty)^2}, \quad 
C_p = \frac{p - p_\infty}{\frac{1}{2}\rho_\infty (M_\infty\times a_\infty)^2}\]
where $L$ and $D$ are the forces in the direction perpendicular and parallel to the freestream $(\infty)$ direction respectively and $a_\infty$ is the speed of sound. We set the following freestream conditions in this work $p_\infty = 28,745~Pa$, $\rho_\infty=0.44 kg/m^3$, $a_\infty=301.86m/s$, $M_\infty=0.734$, $\alpha_\infty=2.79^\circ$.

\subsection{Airfoil shape parametrization}
\label{s:airfoil_shape}
The baseline RAE2822 airfoil shape is parametrized using class shape transformation (CST)~\cite{kulfan2006fundamental, kulfan2008universal} which defines a shape as the product of a \emph{class} function $c()$ and a \emph{shape} function $s()$. The class function $c(\psi)$ is defined as
\begin{equation}
c_{n_1}^{n_2} (\psi) = \psi ^{n_1}(1 - \psi)^{n_2},
\label{e:class_fn}
\end{equation}
where the variable $0 \leq \psi \leq 1$ represents the nondimensional chordwise distance and, $n_1$ and $n_2$ are non-negative real numbers that fully specify the specific class. For instance with $n_1 = 0.5,~n_2 = 1$, $c_{0.5}^{1.0} (\psi) = \sqrt{\psi} (1 - \psi)$ defines airfoils with rounded leading edge and a sharp trailing edge~\cite{kulfan2006fundamental}. The unique shape of an airfoil is then given by the shape function, specified as follows:
\begin{equation}
s(\psi) = \sum_{i=0}^{n} \beta_i \psi^i ,
\label{e:shape_fn1}
\end{equation}
where $\beta_i$ are the undetermined coefficients that are also the shape parameters; we also refer to $\beta_i$ as \emph{CST coefficients} in the rest of the paper. The final shape of the airfoil shape ($z_{cs}$) is then given by
\begin{equation}
z_{cs}(\psi) = c(\psi)s(\psi).
\label{e:CST_final}
\end{equation}
The coefficients $\beta_i,~i=0,\hdots,n_s$ determine the final the airfoil shape given $n_s$, the order of the Bernstein polynomials. An $n_s$th-order CST parametrization has $n_s+1$ parameters; if independent parametrizations are sought for the upper and lower surfaces of the airfoil (which is appropriate for cambered airfoils), then the CST parametrization leads to $2(n_s+1)$ parameters, where the $n_s$ also needs to be determined. Typically, however, $n_s=3-5$ are observed to be adequate to parametrize the airfoil shapes such as the RAE2822 as also observed in other works~\cite{renganathan2018koopman, renganathan2020koopman, renganathan2020machine}. One way to determine $n_s$ and the associated polynomial coefficients is to find the values that minimize some error (e.g., $L_2$) between the true shape of the airfoil and the resulting approximation via CST. In this work, the parameters are determined by solving the following minimization problem, after setting $n_1 = 0.5$ and $n_2=1.0$:

\begin{equation}
\lbrace \beta_i \rbrace_{i=0}^{n_s}, n_s = \underbrace{\text{minimize}}_{n_s, \beta_i, \psi \in [0,1]}~ \left ( z_{true}(\psi) - c(\psi)s(\psi) \right ) ^2 .
\label{e:CST_param}
\end{equation}

In practice, \eqref{e:CST_param} can be simplified as 

\begin{equation}
\lbrace \beta_i \rbrace_{i=0}^{n_s}, n_s = \underbrace{\text{minimize}}_{n_s, \beta_i}~ \left \| \mbf{z}_{true} - \mbf{c} \otimes \mbf{s} \right \|_2 ^2,
\label{e:CST_param_discrete}
\end{equation}
where $\mbf{z}_{true}$, $\mbf{c}$ and $\mbf{s}$ are the discrete approximations of  $z_{true}(\psi)$, $c(\psi)$ and $s(\psi)$, respectively, at $\tilde{\psi} \in \R^{n_s+1}$, which is the discrete approximation of $\psi$ at $n_s+1$ unique points sampled from $\psi$ spanning $[0,1]$, and $\otimes$ is the elementwise product.  In this way, the smallest possible $n_s$ is determined, thereby avoiding overparametrization. Note that $n_s$ is treated as a continuous variable in \eqref{e:CST_param_discrete} and is then rounded off to the nearest integer. Finally, \eqref{e:CST_param_discrete} is solved separately for the upper and lower surfaces of the airfoil.

\begin{figure}
    \centering
    \includegraphics[scale=0.4]{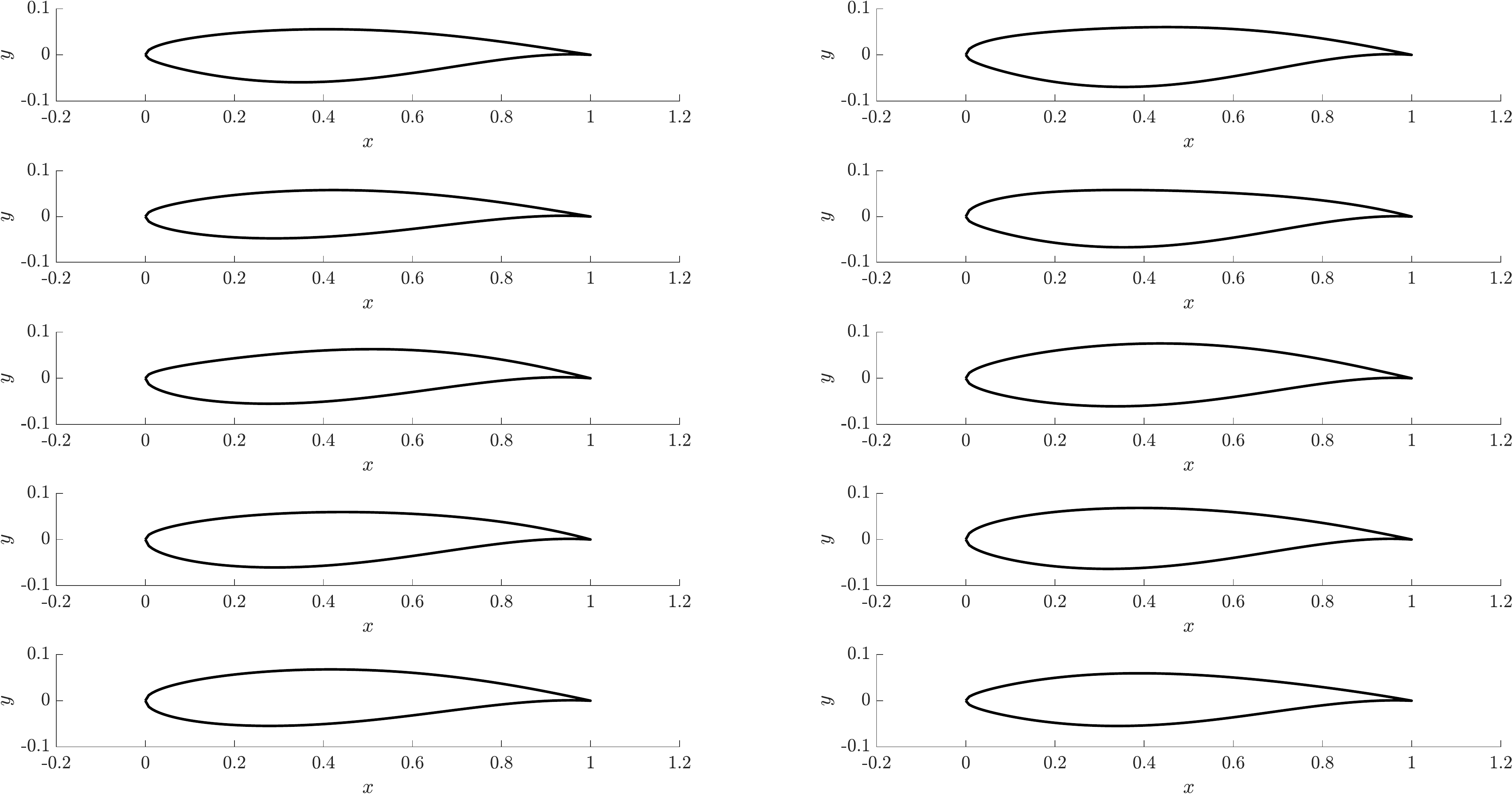}
    \caption{Examples of random airfoils generated by perturbing ($\pm 30\%$) the baseline CST coefficients of RAE2822}
\label{f:RAE2822_CST_Family}
\end{figure}

The RAE2822 airfoil is parametrized by 8 ($n_s = 3$) variables, whose values are denoted $\bs{\beta}_{RAE2822}$ and shown below in \eqref{e:shape_params} which is a concatenation of the CST coefficients for the top and bottom surfaces of the airfoil. Note that \eqref{e:shape_params} represents the \emph{baseline} shape parameters of the RAE2822 airfoil, which are then perturbed to modify the airfoil shape. During ASO, bounds on the parameters are carefully chosen to prevent nonphysical (intersecting) airfoil geometries as well as mesh deformation that could potentially lead to high-fidelity model solver instabilities due to poor-quality elements. In this regard, a domain $\mcl{X}=[\x_L, \x_U]$ is set by perturbing the baseline shape parameters by $\pm 30\%$ to vary the airfoil shape, a sample of which is shown in Figure~\ref{f:RAE2822_CST_Family}. 
\begin{equation}
    \bs{\beta}_{RAE2822} = \begin{bmatrix}
 0.1268 & 0.4670 & 0.5834 & 0.2103 -0.1268 &-0.5425 &-0.5096 & 0.0581
\end{bmatrix}
\label{e:shape_params}
\end{equation}

\subsection{Aerodynamic design optimization}
\label{s:ado}
We address the following optimization problems

\begin{equation}
    \T{\emph{P1:}} \quad \x_* = \underset{\x \in \mcl{X}}{arg\T{min}}~C_d
    \label{e:unconstrained}
\end{equation}
and
\begin{equation}
\begin{split}
   \T{\emph{P2:}} \quad  \x_* = ~&\underset{\x \in \mcl{X}}{arg\T{min}}~C_d \\
    \T{s.t.}\quad & C_l = C^*_l.
\end{split}
\label{e:constrained}
\end{equation}
The domain $\mcl{X}$ is the hyperrectangle $[\x_L, \x_U]$, where
\[\x_L = [0.08876,  0.3269 ,  0.40838,  0.14721, -0.08876, -0.37975, -0.35672,  0.04067]\]
and 
\[\x_U = [0.16484,  0.6071 ,  0.75842,  0.27339, -0.16484, -0.70525, -0.66248,  0.07553].\]
Whereas \eqref{e:unconstrained} is an unconstrained drag minimization subject to the airfoil shape parameters being in a compact space, \eqref{e:constrained} adds a lift constraint, where $C^*_l=1.0$ is the target lift coefficient used in this work. Note that in \eqref{e:constrained}, each evaluation of the constraint is an expensive function call (similar to the objective function) and hence is approximated by an emulator to facilitate cheap evaluations. We now proceed to introduce the optimization approaches used in this work.

\section{Optimization strategies}
\label{s:opt_strats}
\subsection{The adjoint method}
\label{s:Adjoint}
The adjoint method for ASO draws from control theory and treats the boundary shape of the aerodynamic as control parameters. Then the derivatives of the objective function with respect to the control parameters are obtained as a solution of an adjoint equations. The approach is briefly reviewed here but the reader is referred to the following references \cite{jameson1988aerodynamic, jameson1989computational} for further details.

Let $\x \in \mcl{X} \subset \mbb{R}^d$ denote the control parameters that parametrize the shape of the aerodynamic object e.g., aircraft wing or wing section and, $\bs{\omega} \in \mbb{R}^{N}$ denote the vector of state variables of the PDE system that define the flow field, discretized on a grid of size $N$. Then the overall discretized elliptic PDE system---which encodes the boundary conditions---is given as
\begin{equation}
    R(\x, \bs{\omega}) = \mbf{0},
\end{equation}
where  $R:\mbb{R}^d \times \mbb{R}^N \rightarrow \mbb{R}^N$ is the residual operator. Let $J:\mbb{R}^d \times \mbb{R}^N \rightarrow \mbb{R}$ denote the objective function which is derived from the control parameters and the state, that is,
\begin{equation}
    J = J(\x, \bs{\omega}),
\end{equation}
then, the total derivative of the objective function via the chain rule is given by
\begin{equation}
    \f{dJ}{d\x} = \f{\p J}{\p \bs{\omega}} \f{\p \bs{\omega}}{\p \x} + \f{\p J}{\p \x} = J_\omega \omega_\x + J_\x,
    \label{e:dJdx}
\end{equation}
where in \eqref{e:dJdx}, $x_y$ denotes $\f{\p x}{\p y}$. Similarly, the total derivative of $R$ is given by
\begin{equation}
    \f{dR}{d\x} = R_\omega \omega_\x + R_\x = \mbf{0}.
    \label{e:dRdx}
\end{equation}
Note that in \eqref{e:dRdx}, the partial derivatives $R_\omega$ and $R_\x$ can be obtained by differentiating the 
the computer code of the high-fidelity model, e.g., via algorithmic differentiation. Similarly, the cost function $J$, is presumed a \emph{known} function of $\x$ and $\omega$ (since it is computed within the high-fidelity model code), and hence its partial derivatives $J_\omega$ and $J_\x$ can be automatically computed. The $\omega_\x$ is then obtained as the solution of the linear system
\begin{equation}
    R_\omega \bs{\lambda} = -R_\x,
    \label{e:adjoint_eqn}
\end{equation}
where \eqref{e:adjoint_eqn} is called the adjoint equation and $\bs{\lambda} = \f{\p \omega}{\p \x}= [R_\omega]^{-1}[-R_\x]$ . Finally, the total derivative of our cost function is obtained as
\begin{equation}
    \f{dJ}{d\x} = J_\omega[R_\omega]^{-1}[-R_\x] + J_\x
    \label{e:dJdx_final}
\end{equation}
which can then be used within a gradient-based optimizer to find $\x_*$.
Therefore, in summary, the state variable $\omega$ is computed first by solving the high-fidelity or \emph{primal} model. Then, the partial derivatives of the residual $R$ as well as the cost function $J$ with respect to the state variable $\omega$ and the control parameters $\x$ are computed by differentiating the high-fidelity model via e.g., algorithmic differentiation~\cite{griewank2008evaluating,naumann2011art}. Finally, the total derivative of the cost function is computed by solving the adjoint equation for $\bs{\lambda}$ and plugging in $\bs{\lambda}$ in \eqref{e:dJdx_final}. Practically speaking, this means that the gradient of $J$ in terms of an arbitrary number of variables can be computed without the need for additional primal model evaluations done as in, e.g., finite differences. Note that when the gradients obtained via \eqref{e:adjoint_eqn} are used within a gradient-based optimizer, at each step of the optimization search, two expensive function calls are necessary namely, the adjoint solution and the primal solution.

The primal and adjoint solutions are obtained via the commercially available finite volume based code STAR-CCM+. The second order discretization adjoint approach uses the restarted-GMRES driver for improved convergence. A 30 outer iteration limit is imposed and the Krylov subspace dimension is set to 15. The adjoint-based ASO approach is implemented by coupling the high-fidelity primal and adjoint solvers to an external gradient-based optimizer based on the sequential least squares quadratic programming (SLSQP)~(see, e.g., Ch.18 of \cite{nocedal2006numerical} and \cite{kraft1989slsqp}) method. Additionally, the framework uses a control point based free form deformation (FFD) approach to vary the aerodynamic shape. To keep the parametrization tractable, the CST approach (discussed in section \ref{s:airfoil_shape}) is used to generate the shapes and the framework maps the CST coefficient to a higher-dimensional and hence more flexible control point deformation. The overall framework was previously implemented in and successfully validated against benchmark ASO problems in \cite{berguin2018cfd}. The overall method is summarized in Algorithm~\ref{a:adjoint}.

\begin{algorithm2e}[H]
\textbf{Given:} \\
\quad $\mcl{X} = [\x_l, \x_U]$ (domain) \\
\KwResult{$\x_*$}
  \For{$i=1, \ldots, m$, (\T{multistarts}) }{
  Initial guess $\x^{(i)}_0$ sampled uniformly at random from $\mcl{U(\x_L, \x_U)}$ \\
  Compute state variable $\omega$ by solving primal equation\\
  Estimate $\bs{\lambda}$ by solving adjoint equation \\
  Compute gradient $dJ/d\x$ via \eqref{e:dJdx_final} \\
  Find $\x^{(i)}_* = \underset{\x \in \mcl{X}, R(\x, \omega)=0}{arg\T{min}}~ J(\x, \omega)$ (gradient-based minimization)\\
 }
 \quad $\x_* = arg\T{min}_{\x \in \{ \x^{(i)}_*\}}~J(\x, \omega)$
 \caption{Adjoint based optimization}
 \label{a:adjoint}
\end{algorithm2e}

\subsection{Deep neural networks}
\label{s:DNN}
We now present the potential for using differentiable, data-driven maps to replace the expensive primal solution evaluation and dependence on the adjoint solution to compute gradients. For obtaining these maps we use deep neural networks in a supervised learning fashion. In particular, we use the multilayered perceptron (MLP) architecture (a subclass of feedforward artificial neural network) to obtain these maps. A general MLP consists of several neurons arranged in multiple layers. These layers consist of one input and one output layer along with several hidden layers. Each layer (with the exception of an input layer) represents a linear operation followed by a nonlinear activation which allows for great flexibility in representing complicated nonlinear mappings. This may be expressed as 
\begin{align}
    \mathcal{L}^{l}\left(\mathbf{x}^{l-1}\right):=\boldsymbol{w}^{l} \boldsymbol{\x}^{l-1}+\boldsymbol{b}^{l},
\end{align}
where $\mathbf{x}^{l-1}$ is the output of the previous layer, and $\boldsymbol{w}^l,\boldsymbol{b}^l$ are the weights and biases associated with that layer. The output $\mathbf{x}^l$, for each layer except the output layer, is then transformed by a nonlinear activation given by the Tan-Sigmoidal activation i.e.,
\begin{align}
    \phi(x) = \text{tanh} (x) = \frac{e^{x} - e^{-x}}{e^{x} + e^{-x}}.
\end{align}
Note that the activation function of the output layer is the identity function. The inputs $\boldsymbol{\x}^0$, for our experiments, are in $\mathbb{R}^{N_i}$ (i.e., $N_i$ is the number of inputs) and the outputs $\mathbf{x}^K$ are in $\mathbb{R}^{N_0}$ (i.e., $N_o$ is the number of outputs). The final map is given by
\begin{align}
    F: \mathbb{R}^{N_i} \rightarrow \mathbb{R}^{N_o}, \quad \x^0 \mapsto \x^K = \boldsymbol{f}(\x^0 ;(\boldsymbol{w}, \boldsymbol{b})),
\end{align}
where 
\begin{align}
    F(\x^0 ; \boldsymbol{w},\boldsymbol{b})=\left(\mathcal{L}^{K} \circ \phi_{\text{act}} \circ \mathcal{L}^{K-1} \circ \ldots \circ \phi_{\text{act}} \circ \mathcal{L}^{1}\right)(\x^0)
\end{align}
is a complete representation of the neural network and where $\mathbf{w}$ and $\mathbf{b}$ are a collection of all the weights and the biases of the neural network. These weights and biases, lumped together as $\Lambda = \{\mbf{w}, \mbf{b} \}$, are trainable parameters of our map which can be optimized by examples obtained from a training set. The supervised learning framework requires for this set to have examples of inputs in $\mathbb{R}^{N_i}$ and their corresponding outputs $\mathbb{R}^{N_o}$. This is coupled with a cost function $\mathbb{C}$, which is a measure of the error of the prediction of the network and the ground truth. Our cost function is given by
\begin{align}
    \mathbb{C} = \frac{1}{|T|} \sum_{(\x^0, \boldsymbol{y}) \in T}\left\|\boldsymbol{y}-F(\mathbf{x}^0;\Lambda)\right\|^{2}
\end{align}
where our training data set is given by
\begin{align}
    T=\left\{\left(\x_{i}^0, \boldsymbol{y}_{i}\right): \boldsymbol{y}_{i}=f\left(\x_{i}^0\right)\right\}
\end{align}
and where $f\left(\x_{i}^0\right)$ are examples of the true targets obtained via the high-fidelity model evaluations. Gradients of this cost function can then be used in an optimization framework to obtain weights and biases that predict a held-out (i.e., testing) portion of the collected data.

\subsubsection{Coefficient estimation with artificial neural networks}
\label{s:DNN_emul}

The MLP architecture introduced in Section \ref{s:DNN} is used to build a nonlinear mapping between the shape of the airfoil and its force functionals given by the coefficients of lift and drag. Given data for the shape, parameterized by the Bernstein polynomials, and their corresponding lift and drag, the MLP is trained to predict the coefficients of an unseen shape. Therefore, each data point with number of inputs $N_i$, corresponds to samples from an 8-dimensional space, $\mathbb{R}^d$, that controls the shape of the airfoil (i.e., $\mathbf{x}^0 \equiv \mathbf{x}$). The number of outputs $N_o$ is 2 for the lift and drag coefficients respectively; we denote the DNN emulator that predicts the lift and drag coefficients as $F^{C_l}_{DNN}$ and $F^{C_d}_{DNN}$ respectively. 

\begin{figure}
    \centering
    \includegraphics[width=0.95\textwidth]{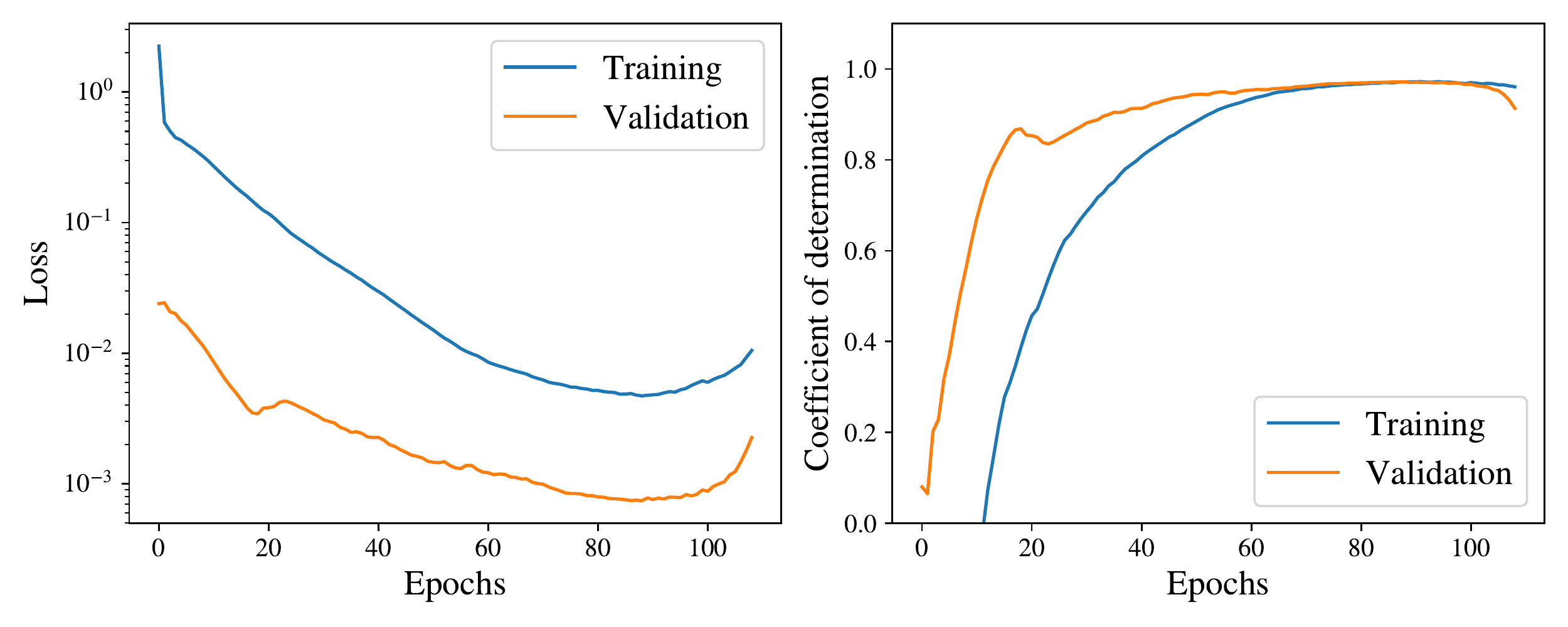}
    \caption{The convergence of the 50 training point MLP showing convergence for both training and validation metrics. The training is terminated once validation losses show a consistent increasing trend of 20 successive epochs without improvement. The consistently smaller validation losses may be due to the significantly smaller size of the validation data set (5 points out of a total of 50).}
    \label{fig:dnn_convergence}
\end{figure}

The available data set is further split into a training set and a validation set with the latter used to assess if the MLP is overfitting on its training observations. The phenomenon of overfitting (or memorization) is often seen in neural network training (particularly for small data sets), where the network performs unreasonably well on the data it observes but fails to generalize to the validation data. Therefore, each MLP training continually requires a monitoring of the validation performance to determine a stopping criterion. The training procedure relies on the Adam stochastic gradient descent algorithm \cite{kingma2014adam}. For each epoch of training (where one epoch denotes a complete observation of the training data), several subsamples (or batches) of the data set are used to compute gradients with respect to the cost function $\mathbb{C}$ which are sequentially used to update the weights and biases of the network. In other words, for each epoch, there are several iterations of the optimization algorithm. The Adam optimizer requires the specification of a learning rate that weights the step-size of the parameter updates which we fix at its default value of 0.001. The convergence of the training may be monitored by tracking both training and validation errors in addition to metrics such as the coefficient of determination (or $R^2$). The validation data also plays a crucial role in the termination of training, where its monitored value is used to determine an early-stopping criteria. In particular, if the validation loss is seen to increase for 20 iterations successively, the training is terminated and the best fit parameters obtain at the lowest validation loss are retained for further deployment. The convergence of our 50 point MLP is shown in Figure \ref{fig:dnn_convergence} indicating a successful optimization. Note that dataset is split into 45 points for training and 5 points for validation. The reader may observe that the validation losses are consistently lower than that of the training, this phenomena may occur when the size of the validation data set is significantly smaller than that of the training.


\begin{figure}
    \centering
    \mbox{
    \begin{subfigure}[htb!]{0.48\textwidth}
        \includegraphics[width=\textwidth]{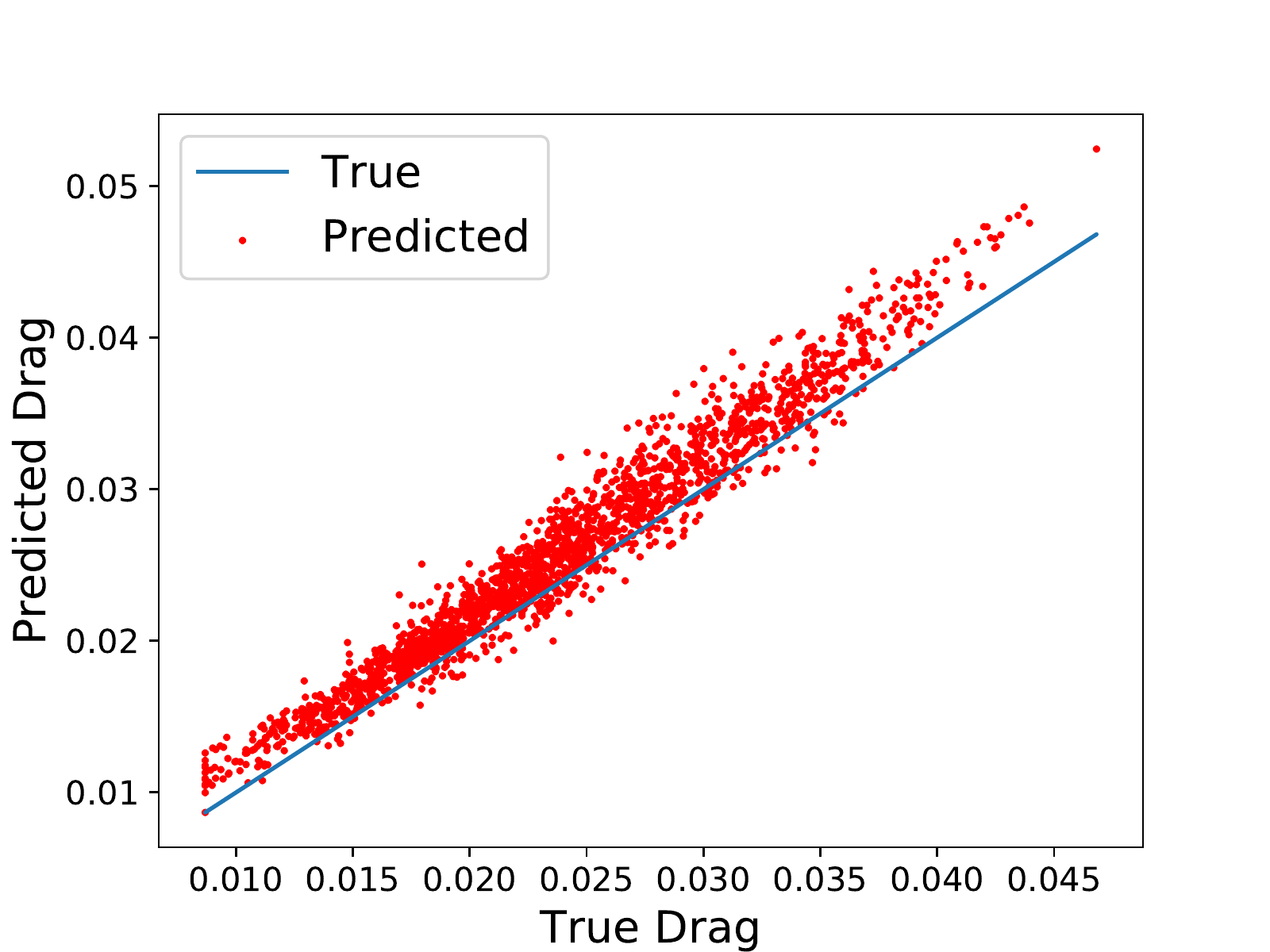}
        \caption{Drag coefficient}
    \end{subfigure}%
    \begin{subfigure}[htb!]{0.48\textwidth}
        \includegraphics[width=\textwidth]{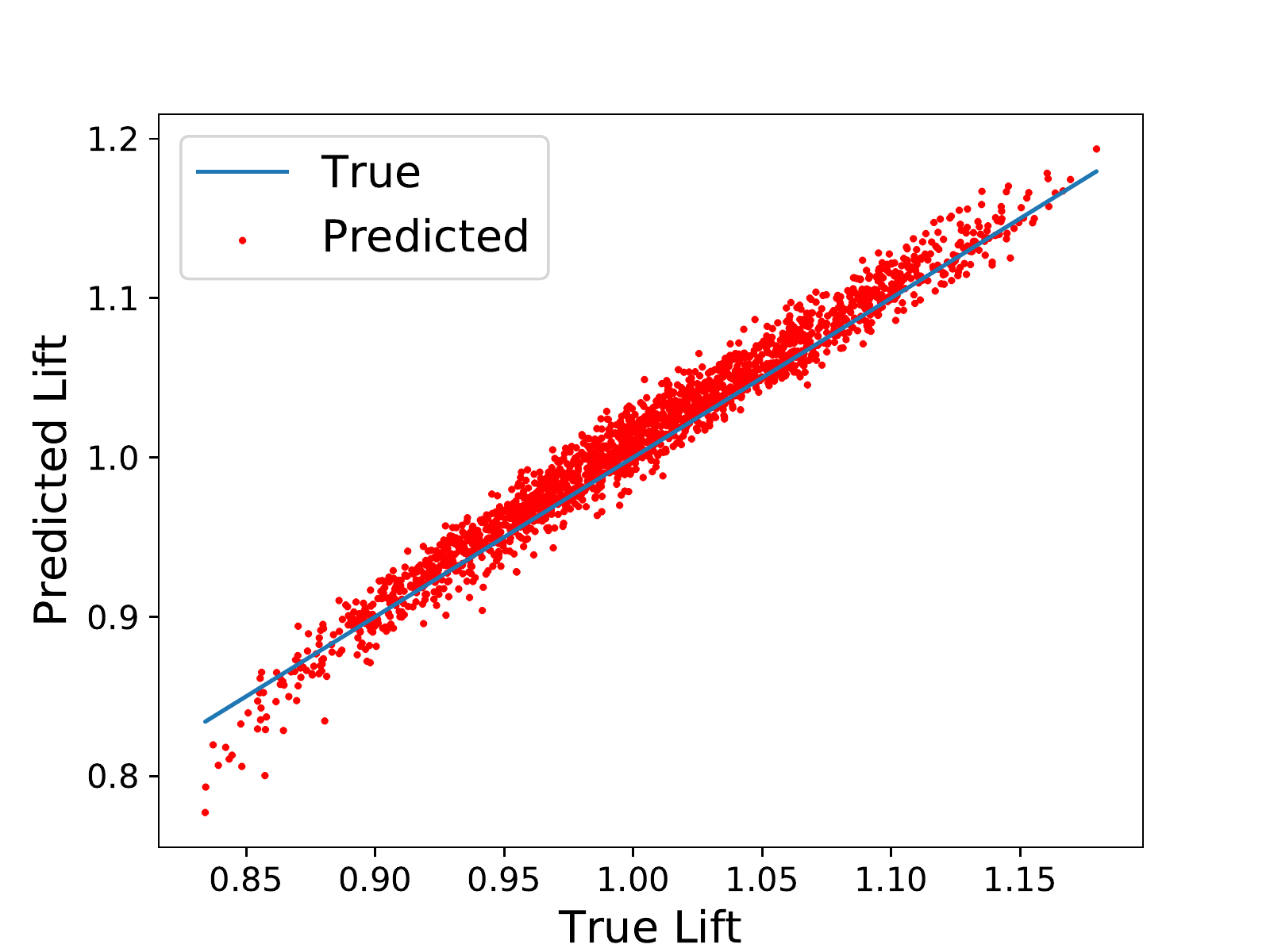}
        \caption{Lift coefficient}
    \end{subfigure}%
    }
    \caption{Test data scatter plots for the true and predicted functionals using the 50 point data set to train our surrogate MLP. The general trends of the outputs with respect to the shape parameters are recovered adequately for the unseen testing data set.}
    \label{fig:dnn_scatter}
\end{figure}

\subsubsection{DNN based optimization}
\label{s:SO}
One of the key advantages of using a neural network for coefficient estimation is the possibility for using analytically differentiable surrogates to compute the gradient with respect to a specified objective function. Indeed, these derivatives are utilized during the training of the neural network when coupled with an objective function that is set to be $\mathbb{C}$. During this phase (known as backpropagation), the gradients of $\mathbb{C}$ with respect to the trainable parameters of the network are used for driving the weights and biases towards a configuration that minimize the error measure.

Once trained, the network may be used to compute gradients of an arbitrary objective function (which depends on $\mathbb{R}^{N_o}$) with respect to an input. Naturally, this may be used for a surrogate adjoint formulation where gradients with respect to the inputs may be used for minimizing this novel objective function. The inputs obtained at the minimal value would then correspond to an optimized configuration in $\mathbb{R}^{d}$. Therefore in this approach, the adjoint calculation is decoupled from a PDE-based system evaluation and is the same cost as a trivial feedforward prediction from the neural network model. Mathematically, our surrogate model relies on an objective function depending on the shape parameters and the trained parameters of the surrogate model and is decoupled from the PDE state variable i.e.,
\begin{align}
    \tilde{J} = \tilde{J}(\mathbf{x};\boldsymbol{w},\boldsymbol{b}).
\end{align}
The gradient of this surrogate adjoint may be computed efficiently by backpropagation through the chain rule of calculus as implemented in most common machine learning packages. In contrast to the training phase of the neural network, backpropagation is now used to obtain the optimal $\mathbf{x}$ for a given objective function $\tilde{J}$. We may summarize this by
\begin{align}
    \nabla_{\mathbf{x}} \tilde{J}=\left(\boldsymbol{w}^{1}\right)^{T} \cdot\left(\phi^{1}\right)^{\prime} \cdots\left(\boldsymbol{w}^{K-1}\right)^{T} \cdot\left(\phi^{K-1}\right)^{\prime} \cdot\left(\boldsymbol{w}^{K}\right)^{T} \cdot\left(\phi^{K}\right)^{\prime} \cdot \nabla_{\boldsymbol{y}} \tilde{J}
\end{align}
where $\phi^{\prime}(x) = \partial \phi / \partial x$. Note that bias terms, $\boldsymbol{b}$, are not treated specially since they correspond to a weight with a fixed input of 1. The general DNN-based optimization approach is summarized in Algorithm~\ref{a:DNNOpt}; note that in the case of constrained optimization, a DNN model for the constraints is first fit with appropriate training examples which is then used in the optimization framework.

\begin{algorithm2e}[H]
\textbf{Given:} \\
\quad $\mcl{D}_n = \lbrace \x_i, y_i \rbrace ~\forall i=1,\hdots,n$, \\
\KwResult{$\x_*$}
\quad Train DNN model to estimate parameters $\Lambda$ \\
  \For{$i=1, \ldots, m$, (\T{multistarts}) }{
  Initial guess $\x^{(i)}_0$ sampled uniformly at random from $\mcl{U(\x_L, \x_U)}$ \\
  Find $\x^{(i)}_* = \underset{\x \in \mcl{X}}{arg\T{min}}~ \tilde{J}(\x;\Lambda)$ (gradient-based $C_d$ minimization)\\
  (Gradients $\nabla_{\mathbf{x}} \tilde{J}$ computed via algorithmic differentiation without dependence on adjoint solver)\\
 }
 \quad $\x_* = arg\T{max}_{\x \in \{ \x^{(i)}_*\}}~\tilde{J}(\x;\Lambda)$
 \caption{DNN based optimization}
 \label{a:DNNOpt}
\end{algorithm2e}

\subsection{Bayesian Optimization}
\label{s:BO}

Bayesian optimization addresses the minimization of a black-box function, i.e. $arg\T{min}_{\x \in \mcl{X} \subset \mbb{R}^d}~f(\x)$. This problem is unsolvable to any theoretical guarantees, unless regularity assumptions are placed on $f(\x)$~\cite{torn1989global}. In this regard, Bayesian optimization (BO)~\cite{brochu2010tutorial} places a Gaussian process (GP) prior on $f(\x) \sim \mcl{GP}(\mu(\x), k(\cdot, \cdot) )$ where, $\mu: \R^d \rightarrow \R$ is the mean function, $k: \R^d \times \R^d \rightarrow \R$ is the covariance function or \emph{kernel} and the GP is fully specified by $\mu$ and $k$. With the assumption of a Gaussian likelihood, the \emph{aposteriori} distribution of $f(\x)$ conditioned on a set of observations $\mcl{D}_n = \lbrace (\x_i, y_i = f(\x_i)), \x_i \in \mcl{X},  \forall i = 1, \ldots, n \rbrace$ can be obtained via Bayes' rule and can be shown to be multivariate normal~\cite{rasmussen:williams:2006}. 

Let $Y(\x)$ denote the GP that approximates the latent (unknown) function $f(\x)$ and $\mbf{y}_n = [y_1,\ldots,y_n]^\top$ be the noise-free observations of $f$ (the high-fidelity model) at $\mbf{X} = [\x_1,\ldots,\x_n ]^\top$, then under a Gaussian likelihood, the posterior distribution is given by
\begin{equation}
    \begin{split}
        Y(\x)|\mcl{D}_n \sim& \mcl{GP}(\mu_n(\x), \sigma_n^2(\x)),\\
        \mu_n(\x) =& \mbf{k}^\top \mbf{K}^{-1} (\mbf{y} - \mu(\mbf{X})) \\
        \sigma_n^2(\x) =& k(\x, \x) - \mbf{k}^\top \mbf{K}^{-1} \mbf{k},
    \end{split}
    \label{e:posterior}
\end{equation}
where $\mbf{K} \subset \mcl{K}$ is the $n\times n$ covariance matrix with $\mbf{K}_{ij} = k(\x_i, \x_j), \forall \x_i, \x_j \in \mcl{X}$ and $\mcl{K}$ being the cone of symmetric positive definite matrices and, $\mbf{k} = [k(\x,\x_1),\ldots,k(\x,\x_n)]^\top$ is an $n\times 1$ vector of covariances. The emulation properties of the GP are driven by the choice of the mean and covariance functions which is explained in the following section.
\subsubsection{Prior assumptions and hyperparameter tuning}
\begin{figure}[ht]
    \centering
    \begin{subfigure}[htb!]{0.32\textwidth}
        \includegraphics[width=\textwidth]{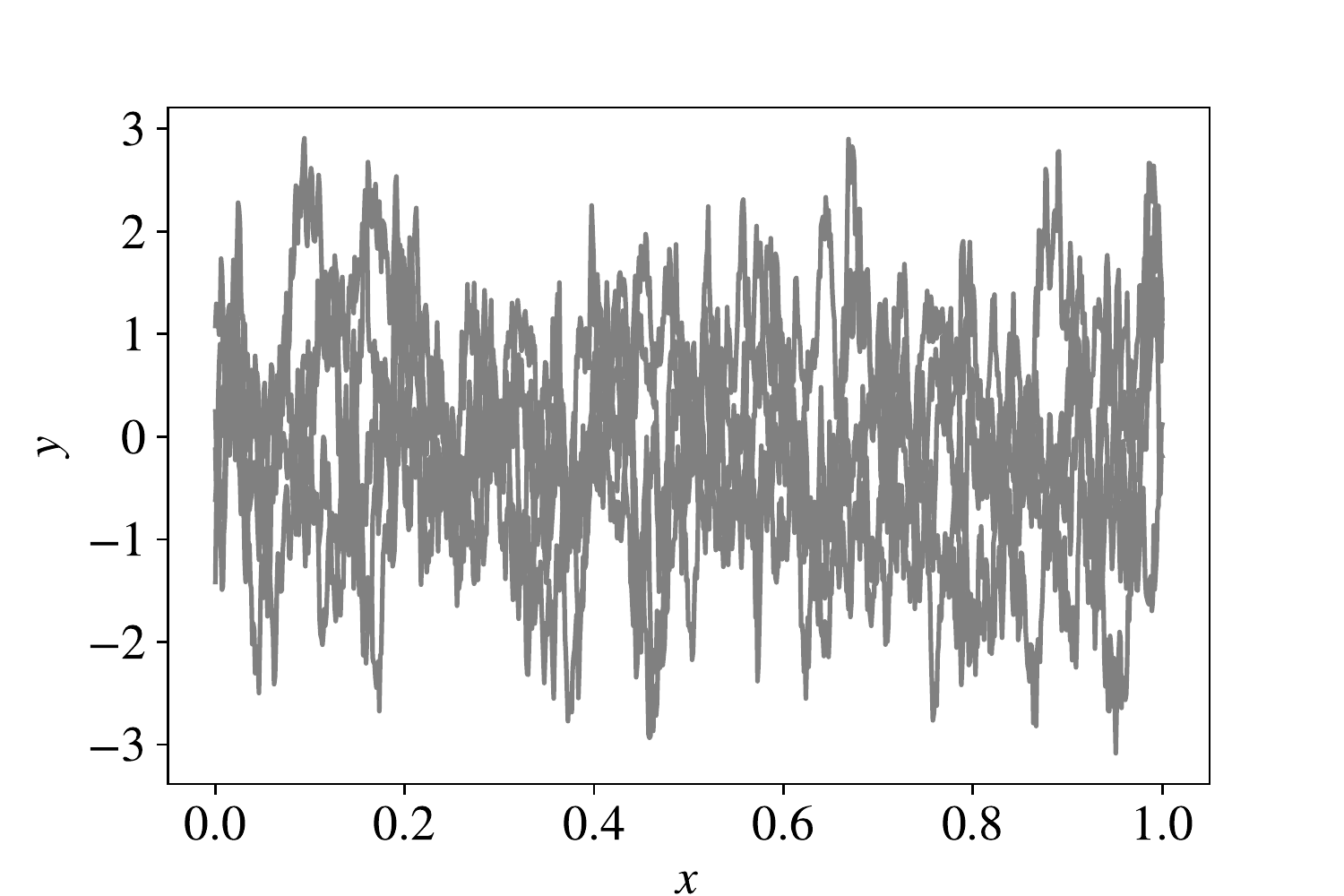}
        \caption{$\nu = 0.5$}
    \end{subfigure}%
    \begin{subfigure}[htb!]{0.32\textwidth}
        \includegraphics[width=\textwidth]{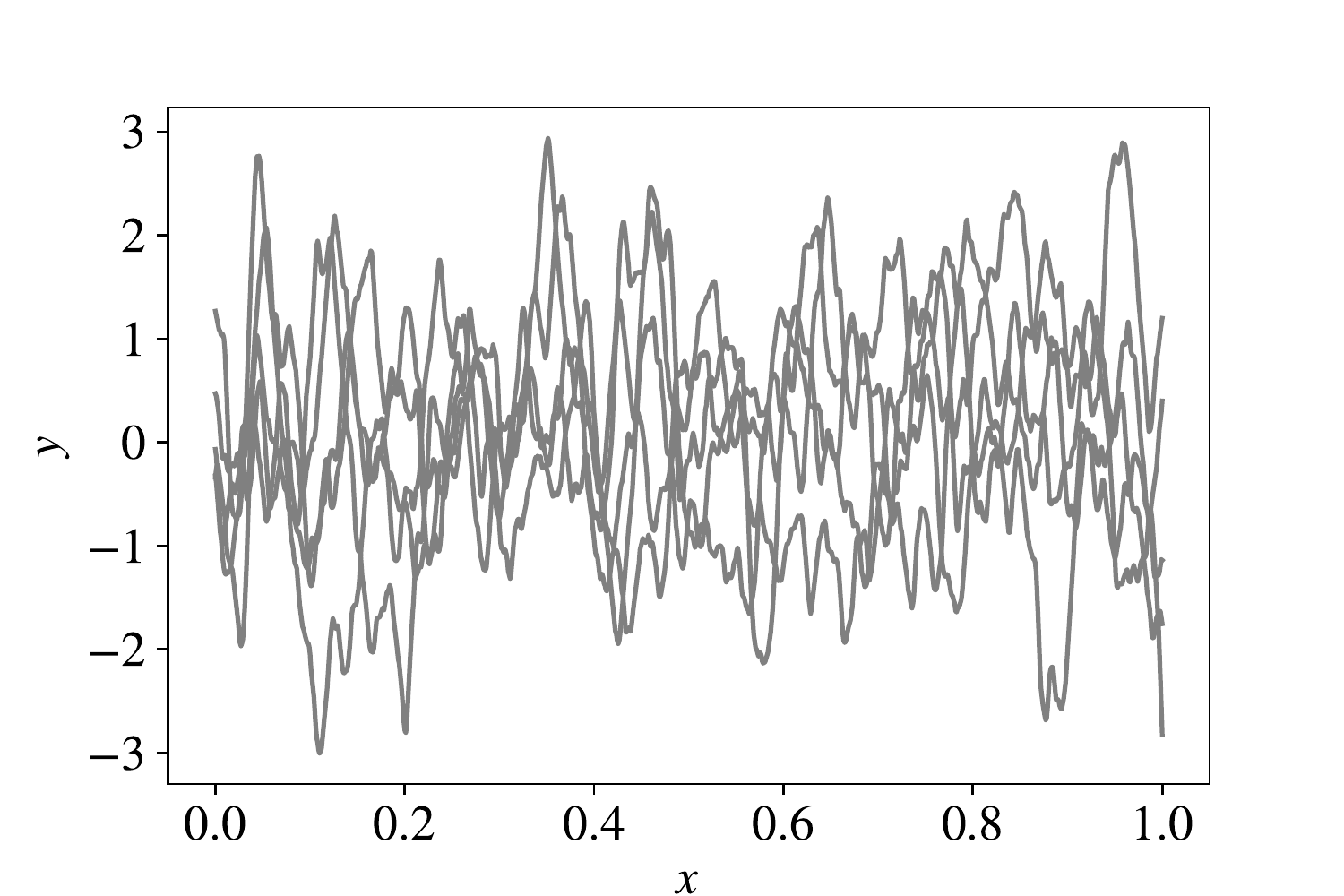}
        \caption{$\nu = 1.5$}
    \end{subfigure}%
    \begin{subfigure}[htb!]{0.32\textwidth}
        \includegraphics[width=\textwidth]{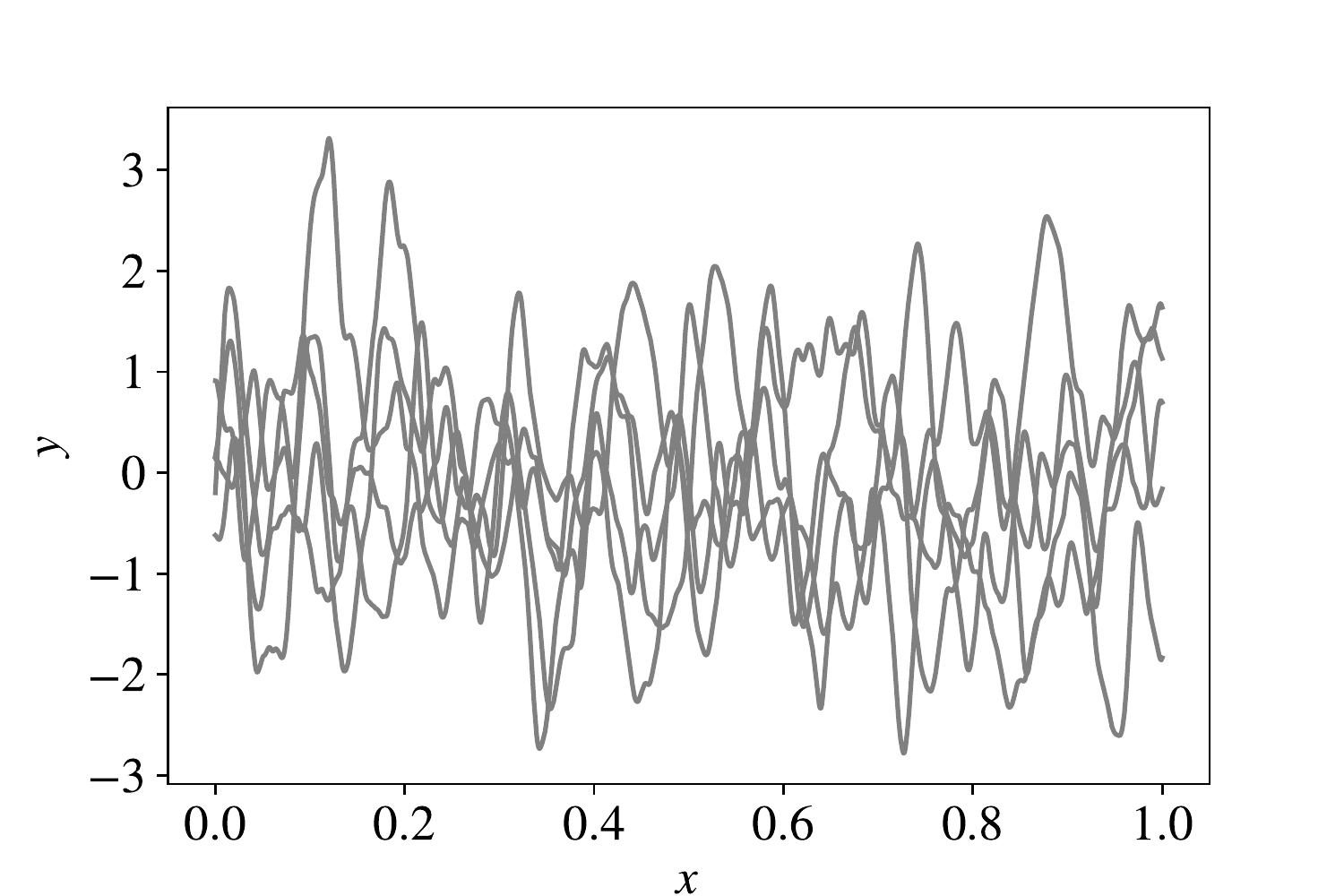}
        \caption{$\nu = 2.5$}
    \end{subfigure}%
    \caption{Realizations from a zero-mean GP with the Matern kernel with $\ell = 0.02$}
    \label{f:matern_realizations}
\end{figure}

The GP prior on $f$ is be specified with parametric functions namely $\mu(\x; \theta_1)$ and $k(\cdot, \cdot; \theta_2)$ whose parameters $\Omega = \{\theta_1, \theta_2 \}$ can be estimated from data. The choice of the mean and covariance functions are subjective and typically driven by available domain knowledge about the problem at hand. In this work, we use a covariance kernel from the Matern class~\cite{stein2012interpolation}, given by 
\begin{equation}
    k(\x, \x') = \f{2^{1-\nu}}{\Gamma(\nu)}\left(\f{\sqrt{2\nu} \|\x-\x' \|}{\ell} \right)^\nu K_{\nu} \left(\f{\sqrt{2\nu} \|\x-\x' \|}{\ell} \right),
\end{equation}
with positive parameters $\nu$ and $\ell$, where $K_{\nu}$ is a modified Bessel function, $\Gamma()$ is the Gamma function and $\| \cdot \|$ denotes the Euclidean distance. The parameter $\nu$ controls the differentiability of the sample paths of the GP and is fixed, whereas $\ell$ is a \emph{lengthscale} parameter that controls the rate of change of the sample paths in $\mcl{X}$ and is a hyperparameter. Figure~\ref{f:matern_realizations} shows sample paths from a zero-mean GP with the Matern class kernel for a few select values of $\nu$; in this work we set $\nu = 3/2$. The GP hyperparameters $\Omega$ are estimated via a maximum likelihood estimation (MLE) procedure frequently followed in fitting GP models~\cite{rasmussen:williams:2006}.

The analytically closed form of the posterior in \eqref{e:posterior} makes GPs attractive to not only emulate an expensive-to-evaluate $f(\x)$, but also efficiently use its emulation and its uncertainty quantification in an optimization setting. BO seeks to use the GP posterior in \eqref{e:posterior} to sequentially choose new points $\x_i~\forall i=n+1,\ldots$, by optimizing a suitably defined \emph{acquisition} function $\alpha(\x)$, in search of the global minimum $f^*$. Because the acquisition function is defined in terms of the GP posterior, it is cheap to evaluate making optimization relatively easy. A generic BO framework is provided in Algorithm~\ref{a:BO}. 

Note that, we assume an overall budget of expensive evaluations $q$ and perform all the BO strategies until the budget runs out and pick the \emph{best} value from all the evaluations. This is in contrast to finite-horizon BO where a dynamic program is solved to optimize a "lookahead" acquisition function; see, e.g., \cite{renganathan2020recursive}.

\begin{algorithm2e}[H]
\textbf{Given:} \\
\quad $\mcl{D}_n = \lbrace \x_i, y_i \rbrace ~\forall i=1,\hdots,n$, \\
\quad $q$ total budget, \\
\quad and GP hyperparameters $\Omega$ \\
\KwResult{$\x_*, y_*$}
  \For{$i=n+1, \ldots, q$, }{
  Target $\xi = \T{min}_{y \in \{y_{1},\ldots,y_{i-1}\}}~y$~\T{(best value)}\\
  Find $\x_i = \underset{\x \in \mcl{X}}{arg\T{max}}~ \alpha_{\T{EI}}(\x)$ (acquisition function)\\
    Observe $y_i$ = $f(\x_i)$\\ 
    Append $\mcl{D}_i = \mcl{D}_{i-1} \cup \lbrace (\x_i, y_i) \rbrace$\\ 
    Update GP hyperparameters $\Omega$ \\
 }
 $y_* = \T{min}_{y \in \{y_{n+1},\ldots,y_q\}}~y$\\
 $\x_* = arg\T{min}_{x \in \{\x_{n+1},\ldots,\x_q\}}~y$
 \caption{Generic Bayesian Optimization}
 \label{a:BO}
\end{algorithm2e}

In this work, we use the expected improvement (EI)~\cite{jones1998efficient} acquisition function which is presented in the following section.

\subsubsection{The expected improvement (EI) acquisition function}
The EI acquisition function defines an improvement function given as
\begin{equation}
  I(\x) = \left(Y(\x)-\xi \right)^+,
    \label{e:improvement}
\end{equation}
where $\xi$ is some user-specified target over which an improvement is sought and $\left(Y(\x)-\xi \right)^+ = \max \left(0, Y(\x)-\xi \right)$. Since $Y(\x)$ is a random variable, so is $I(\x)$ with density $I(\x) \sim \mcl{N}\left((\mu_n (\x) - \xi)^+, \sigma_n^2(\x) \right)$ and so the EI acquisition function $\alpha_{\T{EI}}$ is given as the expectation of $I(\x)$ over the posterior GP $Y(\x)|\mcl{D}_n$
\begin{equation}
    \alpha_\T{EI}(\x) := \mbb{E}_{Y|\mcl{D}_n}(I) = \int_{I=0}^{\infty} I \left\lbrace \f{1}{\sqrt{2\pi} \sigma_n(\x)} \text{exp} \left[- \f{(I - Y(\x) + \xi)^2}{2 \sigma_n^2(\x)} \right]dI \right\rbrace
    \label{e:exp_improvement}
\end{equation}
where the lower limit in the above integral is zero due to the fact that the
improvement is non-negative. Further simplification can be obtained via
transformation of variable and the final acquisition function is given as
\begin{equation}
   \alpha_\T{EI}(\x) =  ( \mu_n(\x) - \xi)  \Phi\left(\f{\mu_n(\x)-\xi}{\sigma_n (\x)}\right) +  \sigma_n(\x) \phi \left( \f{\mu_n(\x) - \xi}{\sigma_n(\x)}\right),
   \label{e:aEI}
\end{equation}
where $\phi()$ and $\Phi()$ denote the standard normal probability density function (PDF) and cumulative density function (CDF) respectively. Whereas the first term in \eqref{e:aEI} favors exploration, the second term favors exploitation~\cite{jones1998efficient, jones2001taxonomy}. However, a key ingredient of the acquisition function to effectively balance exploration and exploitation lies in the accurate emulation of the underlying GP; in other words the accuracy of $\mu_n(\x)$ and $\sigma^2_n(\x)$ in capturing the overall trend and volatility in $f(\x)$. This is typically a challenge in data-scarce situations that arises in the emulation of expensive simulations. To circumvent this, we follow a two-step process where a DNN is first trained the available (small) dataset. Then, the discrepancy in the resulting DNN (due to limited training data) is learned by an interpolating GP; in this regard we specify a zero-mean GP with the Matern class kernel for the covariance function as mentioned previously. In other words, we train the DNN to learn the larger scale trends (or drift) in the data and train a GP model to interpolate the learned model bias; this way what we have is a DNN-enabled GP (DNN-GP) that interpolates the data. This is presented in the following section.

\subsubsection{Deep neural network enhanced expected improvement acquisition function}
\begin{figure}
    \centering
    \includegraphics[width=10cm]{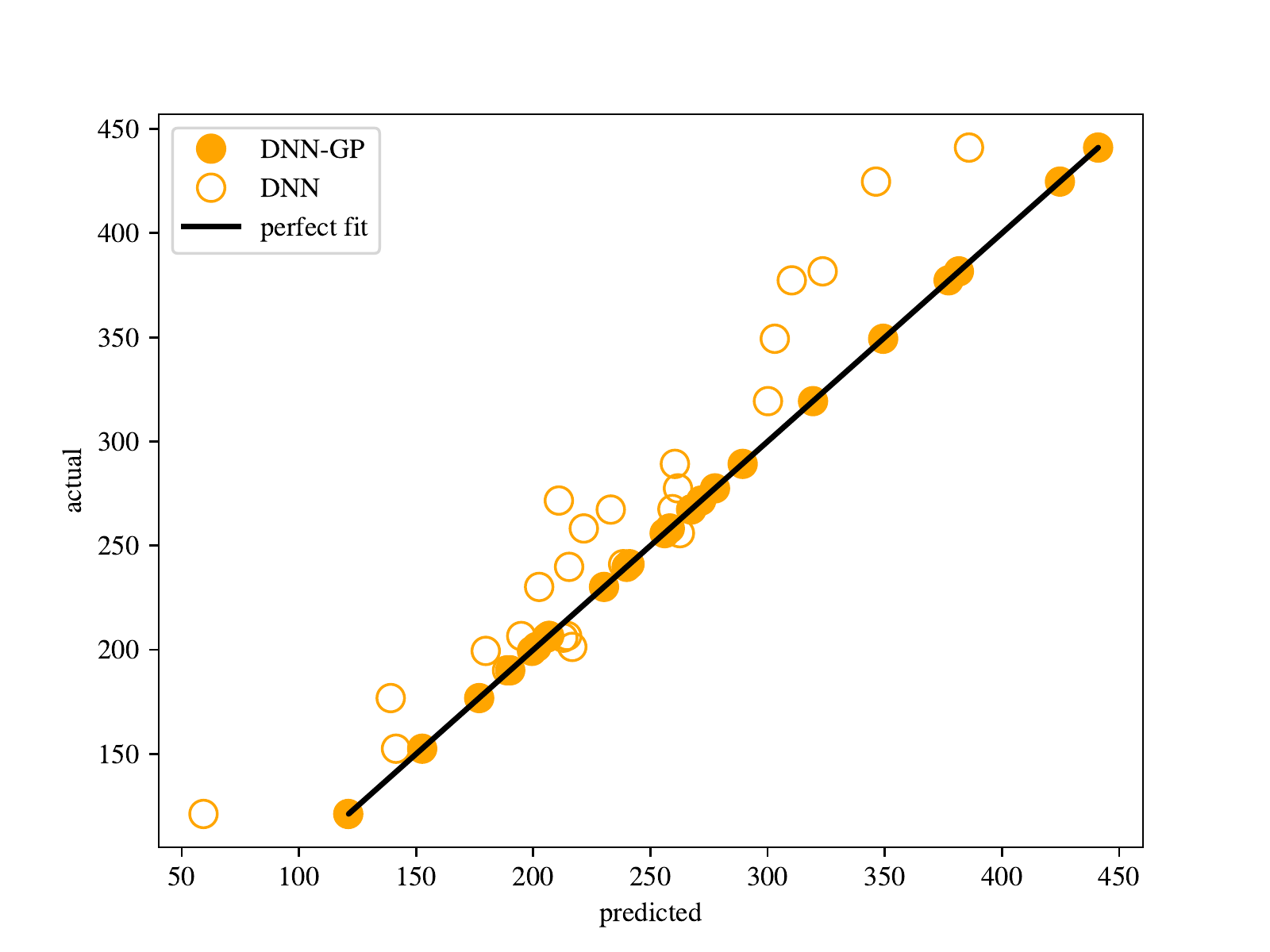}
    \caption{Actual Vs. predicted plot of drag (in counts) on the \emph{training} data. Empty circles are the DNN predictions, filled circles are GP predictions with DNN prior and the straight line shows the \emph{perfect fit} line for reference.}
    \label{f:act_v_pred}
\end{figure}

We now combine the predictive capability of the DNN model introduced in section \ref{s:DNN_emul} and the GP framework that enables efficient black-box optimization by setting 
\begin{equation}
    f(\x) \sim \mcl{GP}(F_{\T{DNN}}(\x), k(\cdot, \cdot)).
\end{equation}
We call this model DNN-GP and recall that $F_{\T{DNN}}: \R^d \rightarrow \R$ emulates $f(\x)$ with a DNN using data $\mcl{D}_n$ given the parameters $\Lambda$. The above equation, given $\Lambda$, can be re-written as
\begin{equation}
    f(\x) - F_{\T{DNN}}(\x)|\Lambda \sim \mcl{GP}(0, k(\cdot, \cdot))
    \label{e:corrected_GP}
\end{equation}
which essentially means that we model the residual between the true function $f(\x)$ and the DNN $F_{\T{DNN}}(\x)$ using a zero-mean GP. The intuition behind \eqref{e:corrected_GP} is that we \emph{calibrate} the DNN model using another probabilistic model which then generalizes the error in the DNN model. We do this to cope with the need to fit a DNN with limited training samples typical of expensive-to-evaluate simulations as considered in this work and hence avoid overfitting. Finally, the (hybrid) DNN-GP is used to construct the EI acquisition function to be used in the BO setting to give
\begin{equation}
   \alpha_\T{DNN-EI}(\x) =  \left( \mu_n(\x) + F_{\T{DNN}}(\x)|\Lambda - \xi\right)~  \Phi\left(\f{\mu_n(\x) + F_{\T{DNN}}(\x)|\Lambda -\xi}{\sigma_n (\x)}\right) +  \sigma_n(\x) ~\phi \left( \f{\mu_n(\x) + F_{\T{DNN}}(\x)|\Lambda +\xi}{\sigma_n(\x)}\right).
   \label{e:aDNN-EI}
\end{equation}
Note that in \eqref{e:aDNN-EI}, one can learn the parameters $\Lambda$ of the DNN and the hyperparameters $\Omega$ jointly by maximizing the marginal likelihood. However, in a small-data setting, this is highly likely to overfit the data due to the difficulties associated with the nonconvex optimization of the marginal likelihood. Furthermore, we want the DNN to explain the overall trends in the data while using the GP to explain only the local volatility which is well facilitated by the two-step procedure that we use. The effect of enhancing the DNN model with the GP is shown in Figure~\ref{f:act_v_pred}, where the DNN-GP interpolates the training data and hence falls along the 45$^\circ$ (perfect fit) line. The DNN-enhanced Bayesian optimization (DNN-BO) procedure is given in Algorithm~\ref{a:DNN-BO}.

\begin{algorithm2e}[H]
\textbf{Given:} \\
\quad Trained DNN model $F_{DNN}$ with parameters $\Lambda$ \\
\quad GP prior mean $\mu = F_{DNN}(\x)$ \\
\quad $\mcl{D}_n = \lbrace \x_i, y_i \rbrace ~\forall i=1,\hdots,n$, \\
\quad $q$ total budget, \\
\quad and GP hyperparameters $\Omega$ \\
\KwResult{($\x_*, y_*)$}
  \For{$i=n+1, \ldots, q$, }{
  
  Find $\x_i = \underset{\x \in \mcl{X}}{arg\T{max}}~ \alpha_\T{DNN-EI}(\x)$ (acquisition function)\\
    Observe $y_i$ = $f(\x_i)$\\ 
    Append $\mcl{D}_i = \mcl{D}_{i-1} \cup \lbrace (\x_i, y_i) \rbrace$\\ 
    Update GP hyperparameters $\Omega$ \\
 }
 \quad $y_* = \T{min}_{y \in \{y_{n+1},\ldots,y_q\}}~y$\\
 \quad $\x_* = arg\T{min}_{x \in \{\x_{n+1},\ldots,\x_q\}}~y$
 \caption{DNN-enhanced Bayesian Optimization (DNN-BO)}
 \label{a:DNN-BO}
\end{algorithm2e}

\subsubsection{DNN-enhanced Bayesian optimization with constraints}
Even though BO is tailored to tackle unconstrained optimization, constraints can be handled with some modifications. We approach the constrained optimization problem defined in \eqref{e:constrained} as described as follows.


We follow the method of Augmented Lagrangian (AL) (see \cite{nocedal2006numerical}, Ch. 17) where we define a modified objective function for \eqref{e:constrained} as 
\begin{equation}
    \underset{\x \in \mcl{X}}{arg\T{min}}~  Q(\x, \lambda) := C_d + \f{\lambda}{2} (C_l - C^*_l)^2,
    \label{e:AL_cons}
\end{equation}
where $\lambda \in \mbb{R}_+$ is the \emph{Lagrange multiplier} whose value must be specified. Intuitively, $\lambda = 0$ entirely removes the penalty and larger values enforce stricter feasibility. In practice, the optimization search is started with moderately large values of $\lambda$ (e.g., 2.0) and progressively increased following a suitable heuristic to adaptively determine the value of $\lambda$ such that $\lambda \rightarrow \infty$ as $i \rightarrow \infty$ where $i$ is the iteration count. 

In the context of BO, $C_d$ and $C_l$ can be modeled by independent GPs ($Y_{C_d}$ and $Y_{C_l}$, respectively) and therefore an approximation of \eqref{e:AL_cons} can be given as
\begin{equation}
    \underset{\x \in \mcl{X}}{arg\T{min}}~  \hat{Q}(\x, \lambda) := Y_{C_d} + \f{\lambda}{2} (Y_{C_l} - C^*_l)^2.
    \label{e:AL_cons_GP}
\end{equation}
However, \eqref{e:AL_cons_GP} does not guarantee that $\hat{Q}$ would still be Gaussian and therefore an approximation for $\hat{Q}$ is required via Monte Carlo sampling because EI is not available in closed form. One can then use a stochastic optimization approach such as sample average approximation (SAA)~\cite{spall2005introduction}.

Instead of working with the non-Gaussian model of the objective function in \eqref{e:AL_cons_GP}, we choose to model the AL using a single GP, an approach similar in spirit to \cite{gramacy2016modeling}. That is, we set
\begin{equation}
    \begin{split}
        Q(\x, \lambda) \sim & \mcl{GP}(\mu_{DNN}(\x), k(\cdot, \cdot)), \\
        \mu_{DNN}(\x) = & F^{C_d}_{DNN}(\x) + \f{\lambda}{2} (F^{C_l}_{DNN}(\x) - C^*_l)^2,
    \end{split}
    \label{e:ALGP}
\end{equation}
where the observations for the GP are $\{ y_i \} = \{C^{(i)}_d + \f{\lambda}{2} (C^{(i)}_l - C^*_l)^2 \}, ~\forall i = 1, \ldots, n$ and, $F^{C_d}_{DNN}(\x)$ and $F^{C_l}_{DNN}(\x)$ are the DNN emulators of $C_d$ and $C_l$ respectively. In this regard, we set a fixed value of $\lambda = 25$ throughout the optimization process. The overall method for the DNN-enhanced BO approach is summarized in Algorithm~\ref{a:DNN-BO-cons}.

\begin{algorithm2e}[H]
\textbf{Given:} \\
\quad Trained DNN model $F_{DNN}(\x)$ \\
\quad Lagrange multiplier $\lambda$ \\
\quad $\mcl{D}_n = \lbrace \x_i, y_i \rbrace ~\forall i=1,\hdots,n$, (Augmented Lagrangian observations)\\
\quad $q$ total budget, \\
\quad and GP hyperparameters $\Omega$ \\
\KwResult{$(\x_*, y_*)$}
  \For{$\ell=n+1, \ldots, q$, }{
  
  Find $\x_i = \underset{\x \in \mcl{X}}{arg\T{max}}~ \alpha(\x)$ (acquisition function)\\
    Observe $y_i$ = $f(\x_i)$\\ 
    Append $\mcl{D}_i = \mcl{D}_{i-1} \cup \lbrace (\x_i, y_i) \rbrace$\\ 
    Update GP hyperparameters $\Omega$ \\
 }
 \quad $y_* = \T{min}_{y \in \{y_{n+1},\ldots,y_q\}}~y$\\
 \quad $\x_* = arg\T{min}_{\x \in \{\x_{n+1},\ldots,\x_q\}}~y$
 \caption{Constrained DNN-enhanced Bayesian Optimization (DNN-BOc)}
 \label{a:DNN-BO-cons}
\end{algorithm2e}

\section{Results and discussion}
\label{s:results}
The proposed DNN-based and DNN-GP-based (DNN-BO) methods are demonstrated on problems \emph{P1} and \emph{P2} defined previously namely, the unconstrained and $C_l$-constrained $C_d$ minimization, both of which search for the optimum in a compact domain $\mcl{X}$. The DNN model is trained with $n=50$ points distributed uniformly random in $\mcl{X}$ where 45 points are used for training and 5 for validation. The same training points used to fit the DNN are reused to fit the DNN-GP (i.e., the DNN-enhanced GP) model. The adjoint-based optimization results are computed via Algorithm~\ref{a:adjoint} with the SLSQP optimizer, where the optimization is run for p=50 iterations with a tolerance on $\x$, $\tau_{\x} = 1e-10$ and tolerance on $C_d$, $\tau_{C_d} = 1e-5$. Additionally, the algorithm is repeated for $m=6$ multistarts and the best (with lowest objective value) optimum is presented. Therefore, the adjoint approach uses $p\times2\times m$ high-fidelity evaluations compared to the $n=50$ evaluations the DNN and DNN-BO approaches use. The DNN-BO method is started with the 40 training samples which are a subset of the training samples used by the DNN and then run for and additional 10 iterations, thereby keeping the total number of high-fidelity evaluations consistent at 50 between the DNN and DNN-BO (and DNN-BOC) algorithms. Note that for the surrogate-based approaches, the multistarts are done on the optimization of a cheap-to-evaluate function and hence the computational cost is relatively negligible.

\begin{figure}[ht]
    \centering
    \begin{subfigure}{0.5\textwidth}
     \includegraphics[width=\linewidth]{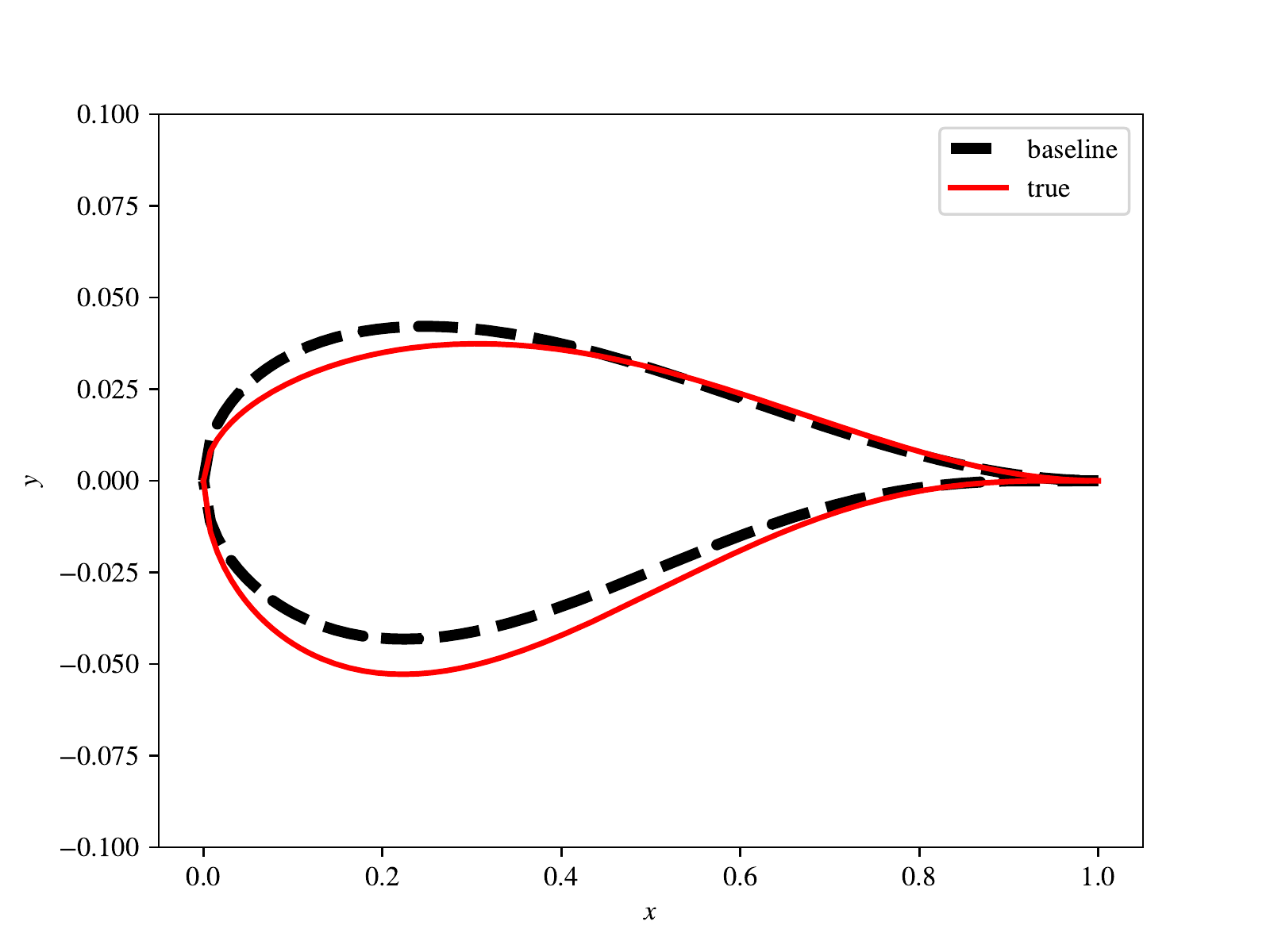}
     \caption{\emph{P1}: Airfoil shapes}
     \end{subfigure}%
    \begin{subfigure}{0.5\textwidth}
     \includegraphics[width=\linewidth]{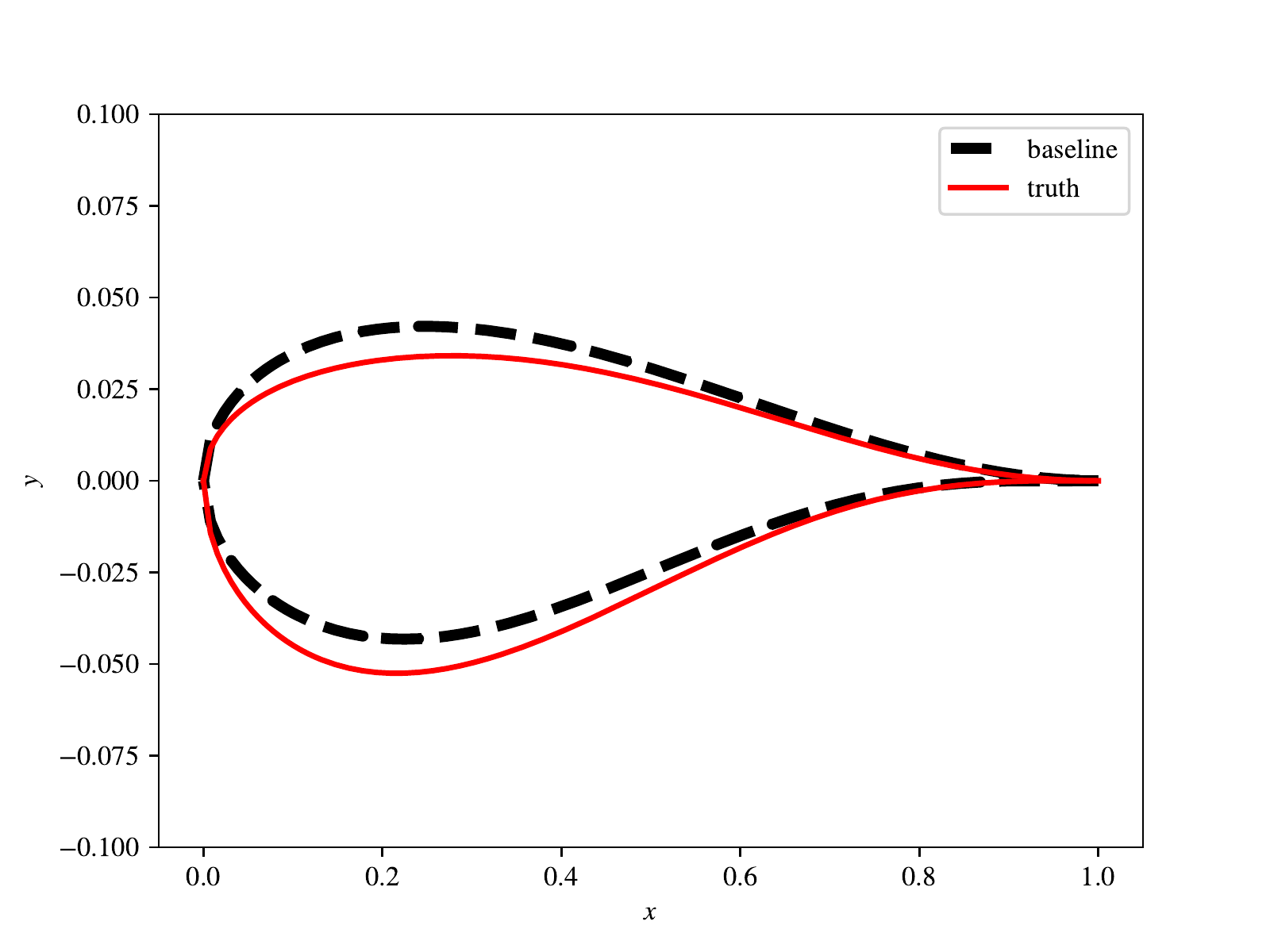}
     \caption{\emph{P2}: Airfoil shapes}
     \end{subfigure}\\
    \begin{subfigure}{0.5\textwidth}
     \includegraphics[width=\linewidth]{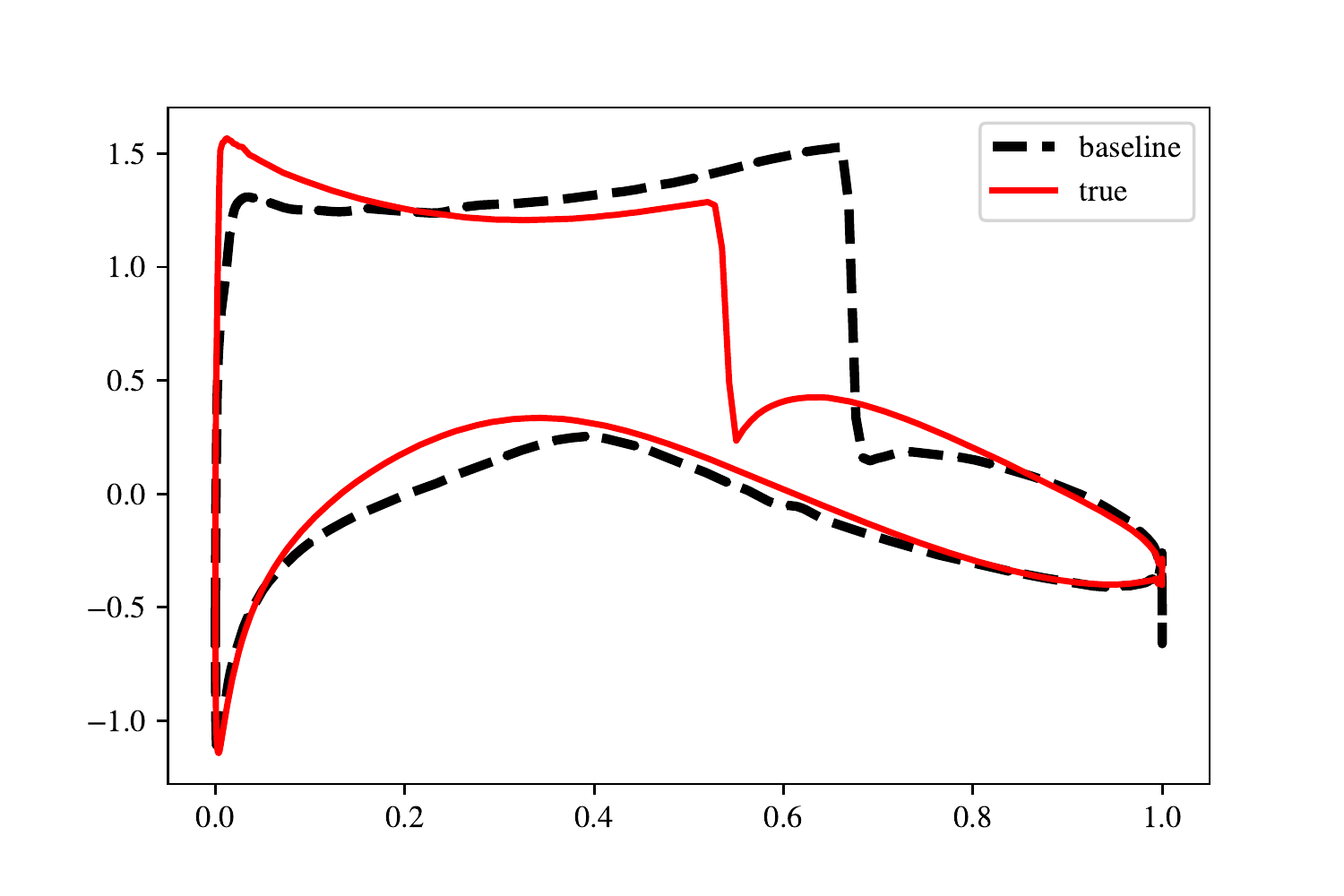}
     \caption{\emph{P1}: $C_p$ distribution}
     \end{subfigure}%
    \begin{subfigure}{0.5\textwidth}
     \includegraphics[width=\linewidth]{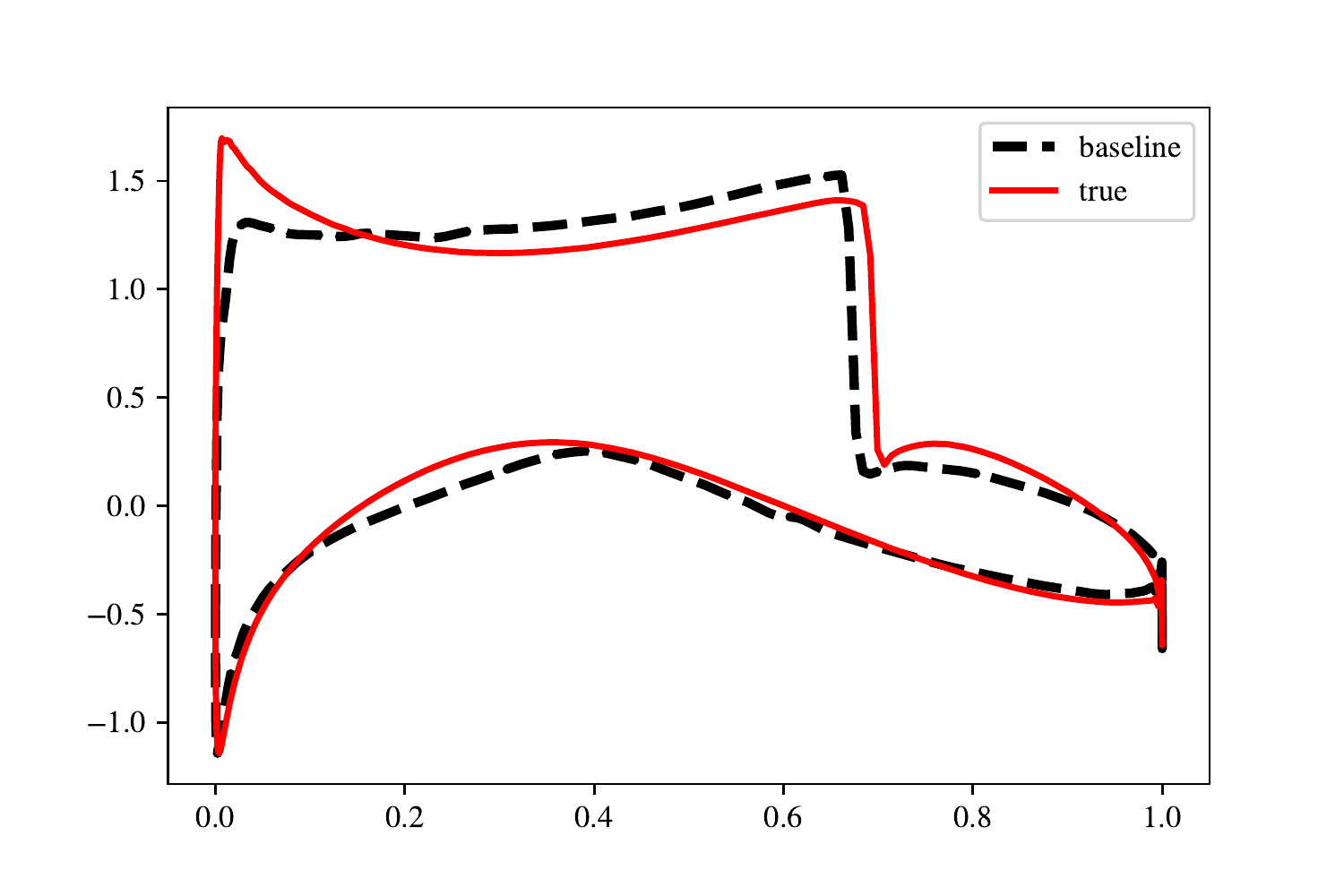}
     \caption{\emph{P2}: $C_p$ distribution}
     \end{subfigure}\\     
    \caption{Baseline and optimized airfoil shapes and $C_p$ distribution}
    \label{f:base_opt}
\end{figure}

To assess the predictive accuracy of all the methods, a dense set of 2000 points in $\mcl{X}$ is generated via Latin Hypercube sampling and the high-fidelity solver is evaluated; then the best shapes in terms of the objectives in \emph{P1} and \emph{P2} are selected from the set. The baseline RAE2822 and the true optimized shapes are compared in Figure~\ref{f:base_opt} along with their $C_p$ distributions.

\subsection{\emph{P1}: unconstrained $C_d$ minimization}
\label{s:uncon_min}
\begin{figure}[ht]
    \centering
    \begin{subfigure}{0.5\textwidth}
     \includegraphics[width=\linewidth]{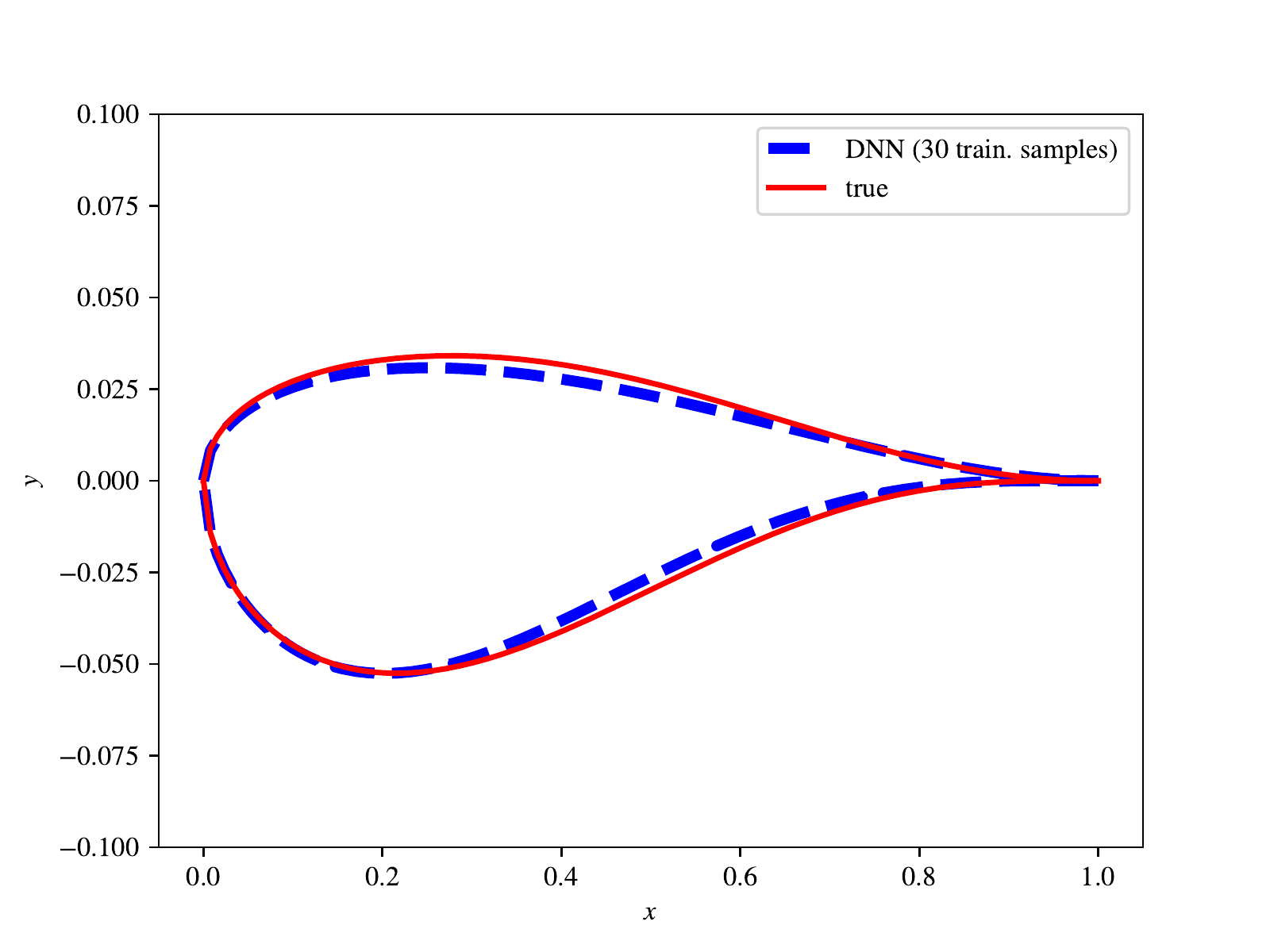}
     \caption{DNN-based optimization (30 training samples)}
     \end{subfigure}%
    \begin{subfigure}{0.5\textwidth}
     \includegraphics[width=\linewidth]{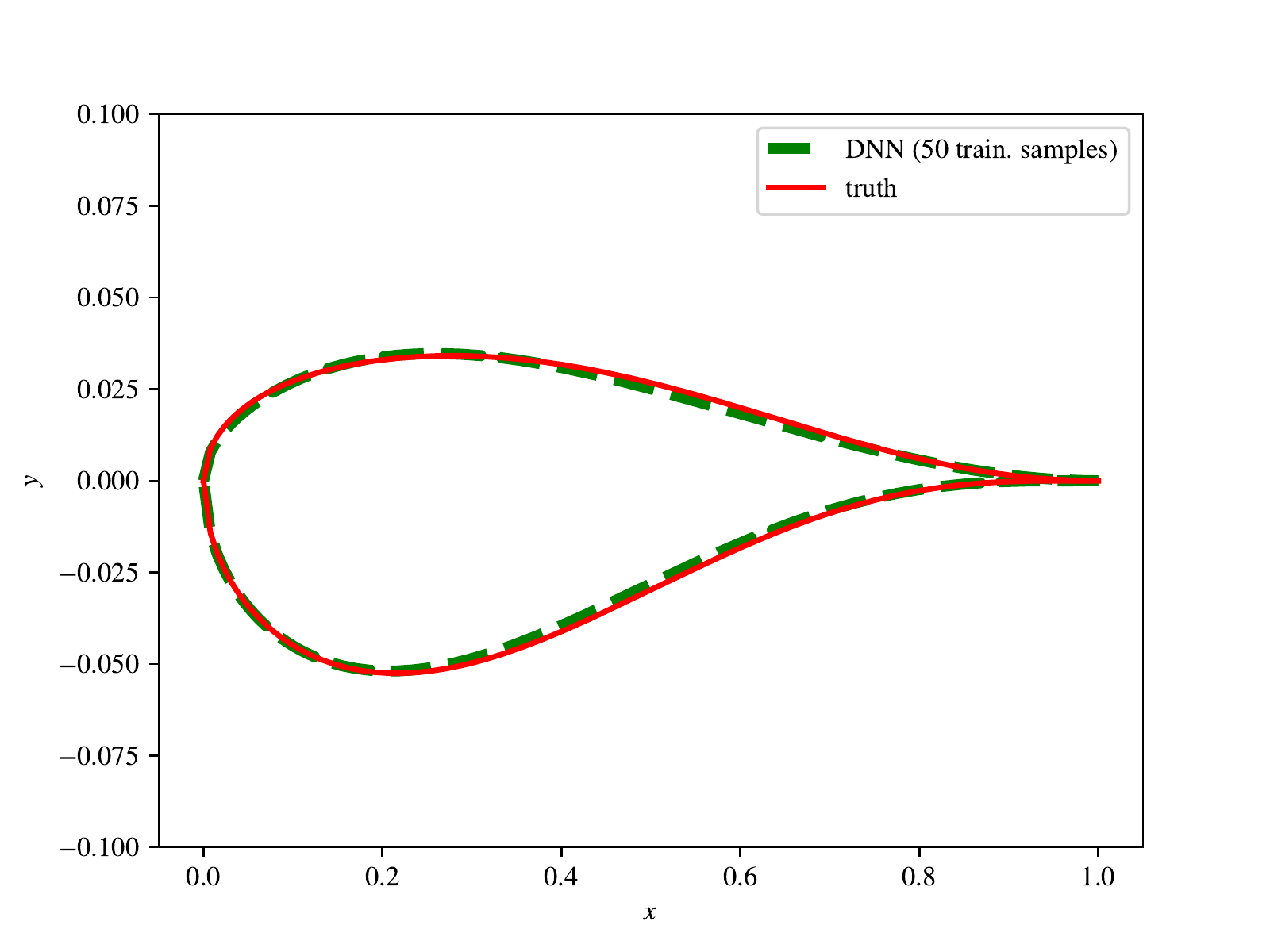}
     \caption{DNN-based optimization (50 training samples)}
     \end{subfigure}\\
    \begin{subfigure}{0.5\textwidth}
     \includegraphics[width=\linewidth]{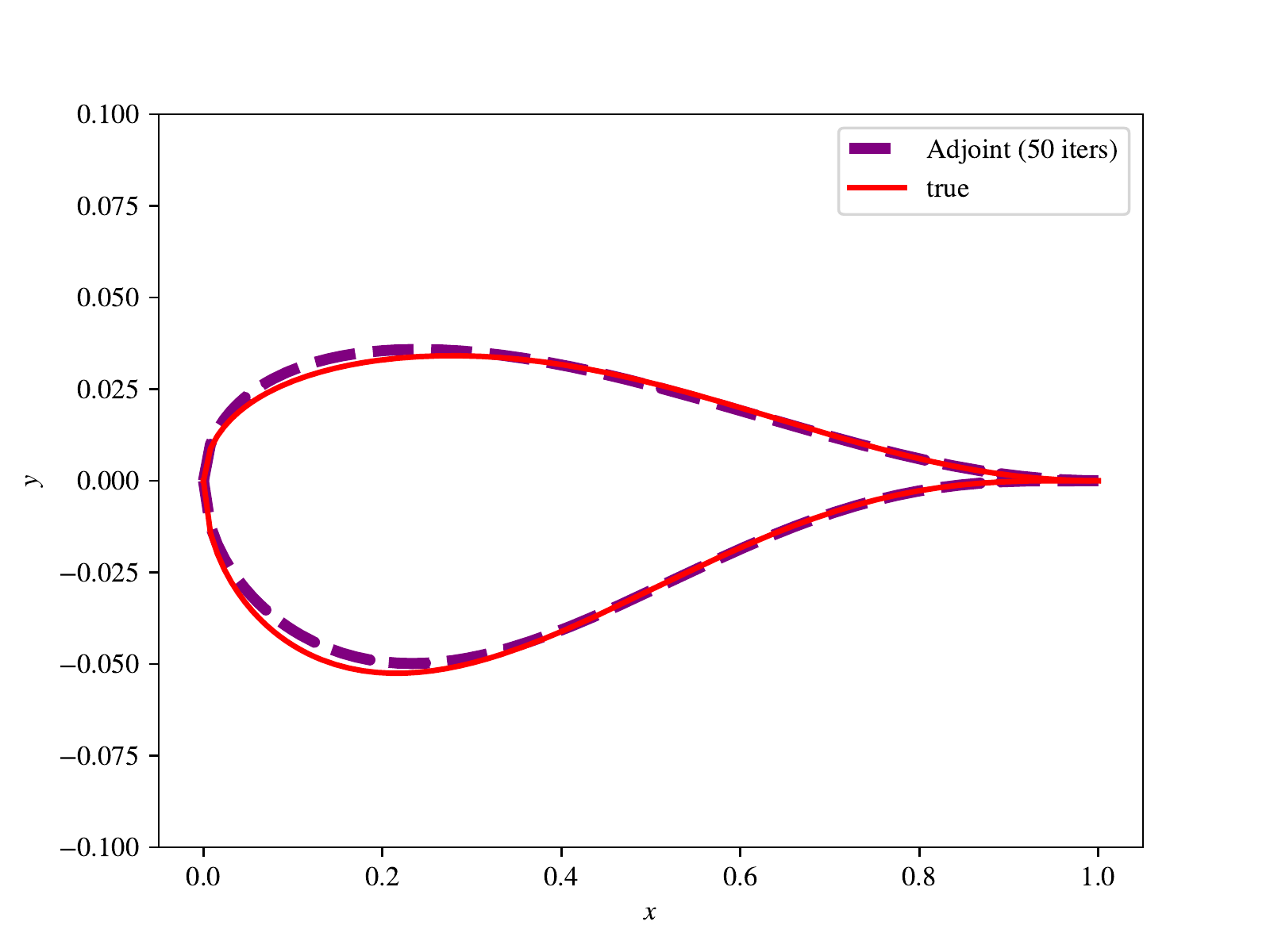}
     \caption{Adjoint-based optimization ($p=50$, $m=6$)}
     \end{subfigure}%
    \begin{subfigure}{0.5\textwidth}
     \includegraphics[width=\linewidth]{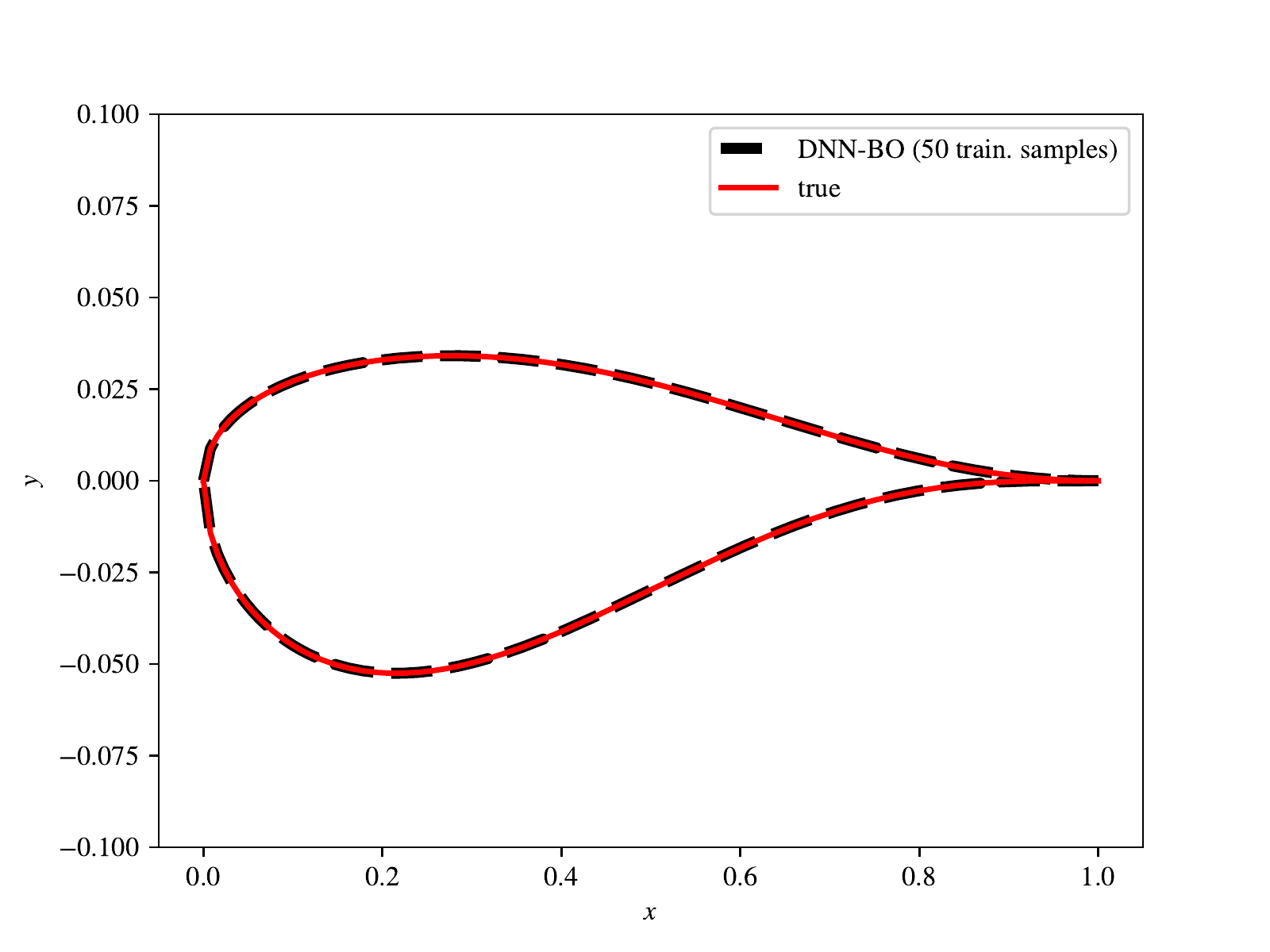}
     \caption{DNN-enhanced Bayesian optimization (50 training samples)}
     \end{subfigure}     
    \caption{Airfoil shapes for \emph{P1:} Unconstrained $C_d$ minimization}
    \label{f:uncon}
\end{figure}

The airfoil shape results of the unconstrained minimization are shown in Figure~\ref{f:uncon}. In the top row, the DNN-based optimization is shown, where the result of a DNN trained with 30 training points is also included to show the impact of the training dataset size on the predictive accuracy. Between 30 and 50 training points, the predictive accuracy drastically improves primarily due to an improved DNN model that is trained with 50 points. However, the DNN-BO (bottom right) further improves the predictive accuracy where the predicted airfoil shape almost exactly matches the true shape. We attribute the success of the DNN-BO primarily to (i) the highly informative prior knowledge specified via the DNN model, (ii) the interpolatory nature of the resulting DNN-GP model and (iii) the additional flexibility that the GP offers to learn the local volatility in the data that the DNN fails to capture with a rather small training set. 

The adjoint result is shown in the bottom left, where the predicted optimum shape, while very close, is not as accurate as the DNN-BO prediction. Note that while the result shown is after 50 iterations of the adjoint-based optimizer multistarted 6 times, however, we anticipate that with more iterations and multistarts, the adjoint results could potentially converge to the true shape. On the other hand, we used the result after 50 iterations to enable a fair comparison with the surrogate-based approaches.

\begin{figure}[ht]
    \centering
    \begin{subfigure}{0.5\textwidth}
     \includegraphics[width=\linewidth]{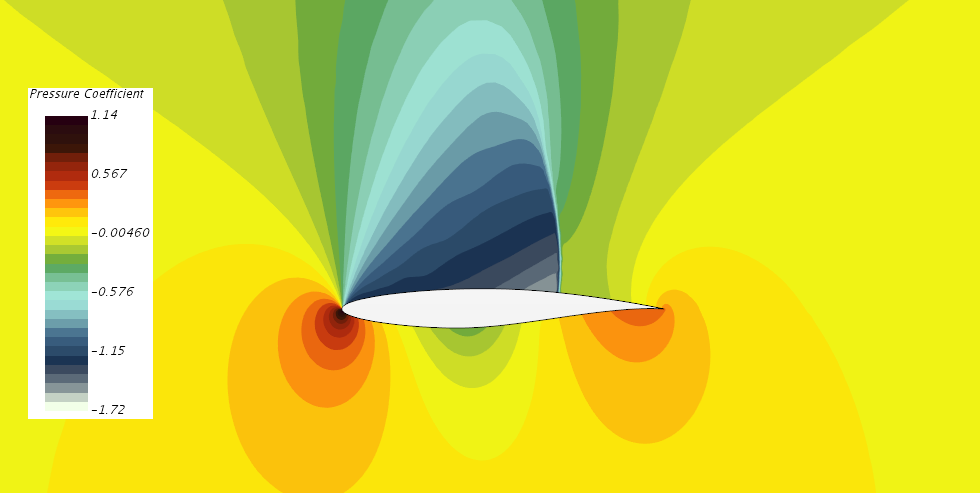}
     \caption{Baseline}
     \end{subfigure}%
    \begin{subfigure}{0.5\textwidth}
     \includegraphics[width=\linewidth]{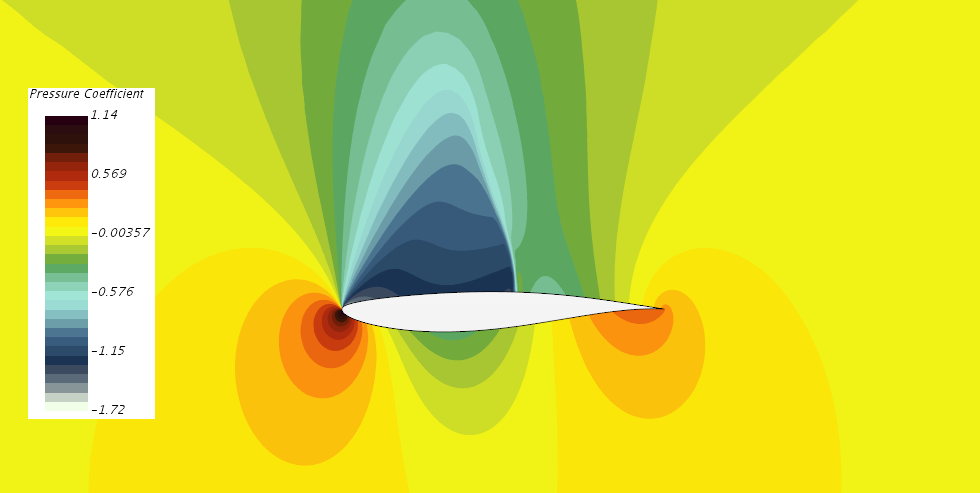}
     \caption{True optimized}
     \end{subfigure}\\
     \begin{subfigure}{0.5\textwidth}
     \includegraphics[width=\linewidth]{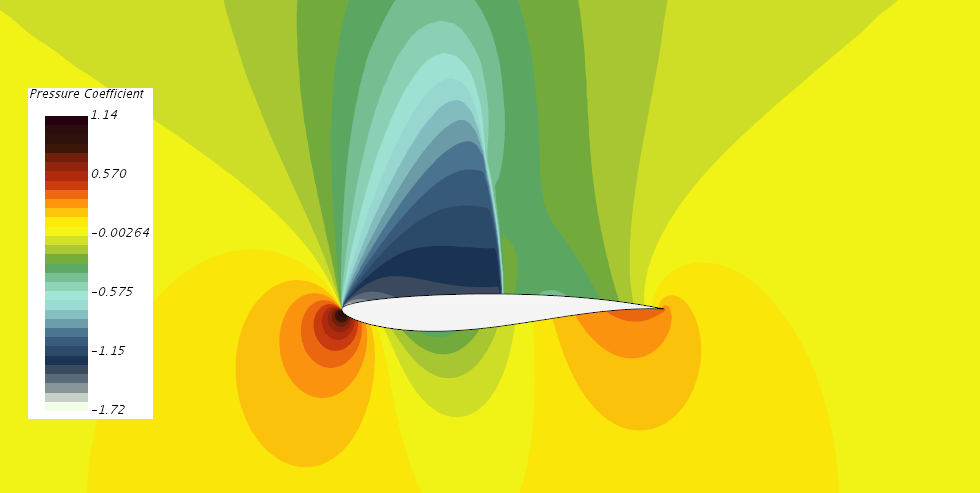}
     \caption{DNN optimized}
     \end{subfigure}%
    \begin{subfigure}{0.5\textwidth}
     \includegraphics[width=\linewidth]{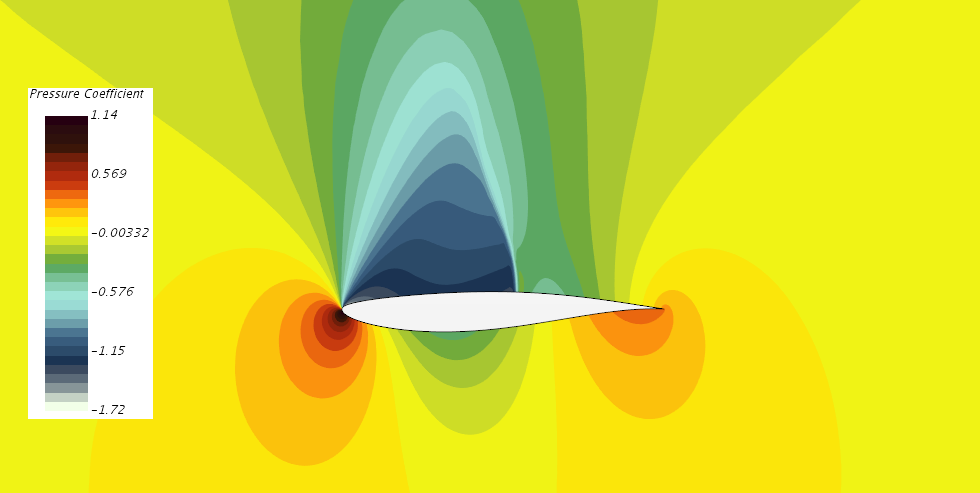}
     \caption{DNN-BO optimized}
     \end{subfigure}\\
    \caption{Pressure distributions  for \emph{P1}: unconstrained $C_d$ minimization}
    \label{f:uncon_pressure}
\end{figure}

\begin{figure}[ht]
    \centering
    \begin{subfigure}{0.5\textwidth}
     \includegraphics[width=\linewidth]{uncon_true_base_cp.pdf}
     \caption{Baseline}
     \end{subfigure}%
    \begin{subfigure}{0.5\textwidth}
     \includegraphics[width=\linewidth]{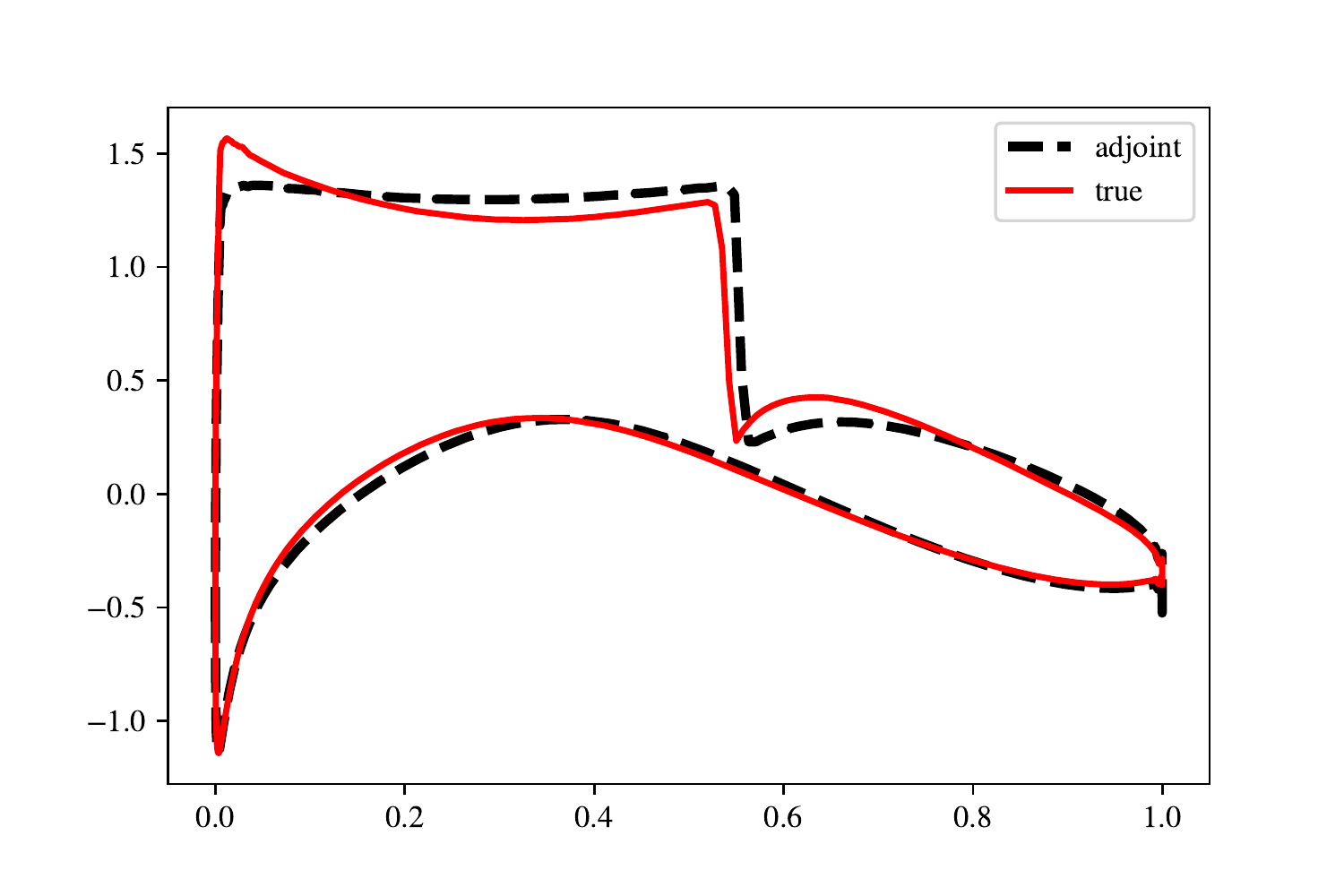}
     \caption{Adjoint}
     \end{subfigure}\\
     \begin{subfigure}{0.5\textwidth}
     \includegraphics[width=\linewidth]{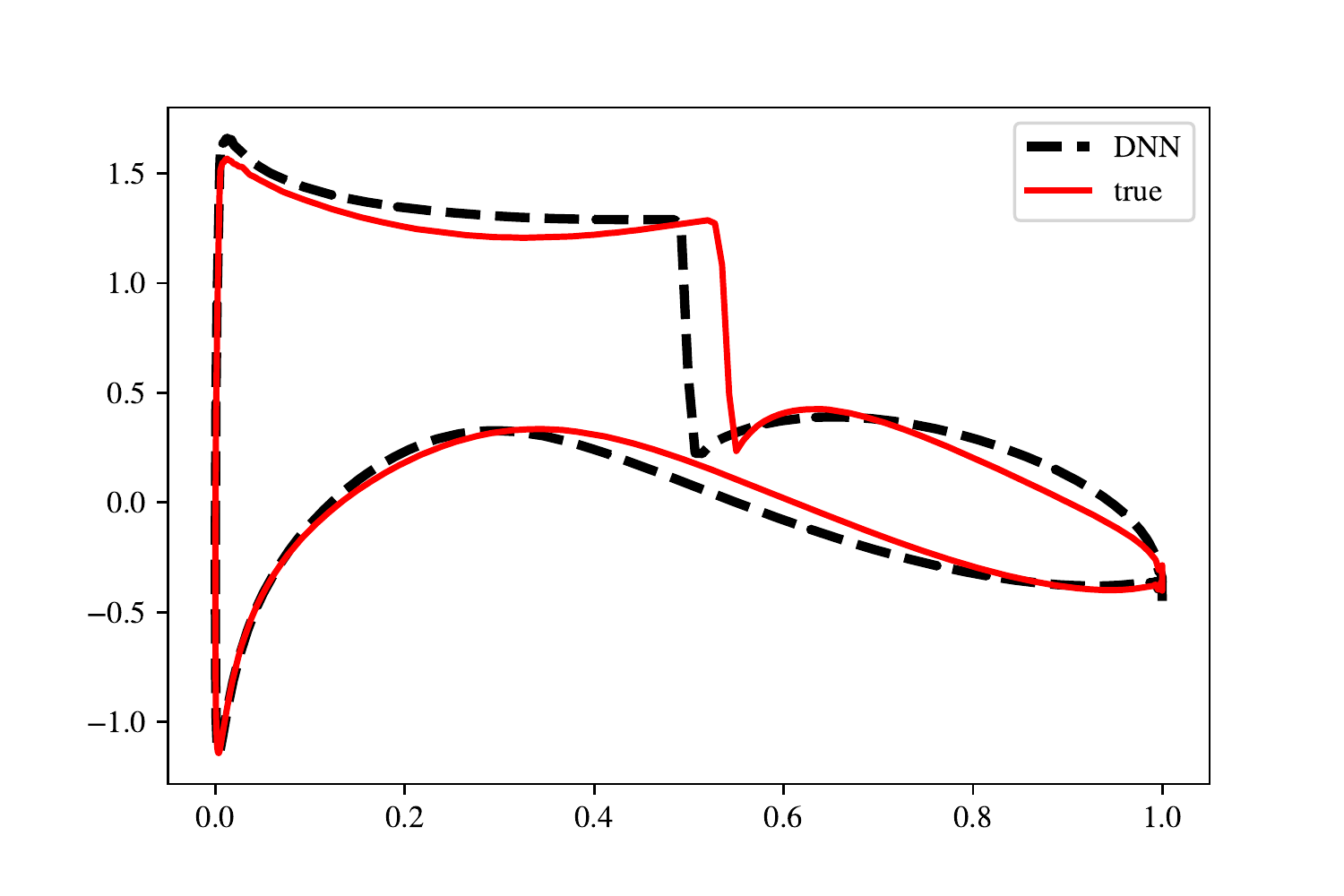}
     \caption{DNN optimized}
     \end{subfigure}%
    \begin{subfigure}{0.5\textwidth}
     \includegraphics[width=\linewidth]{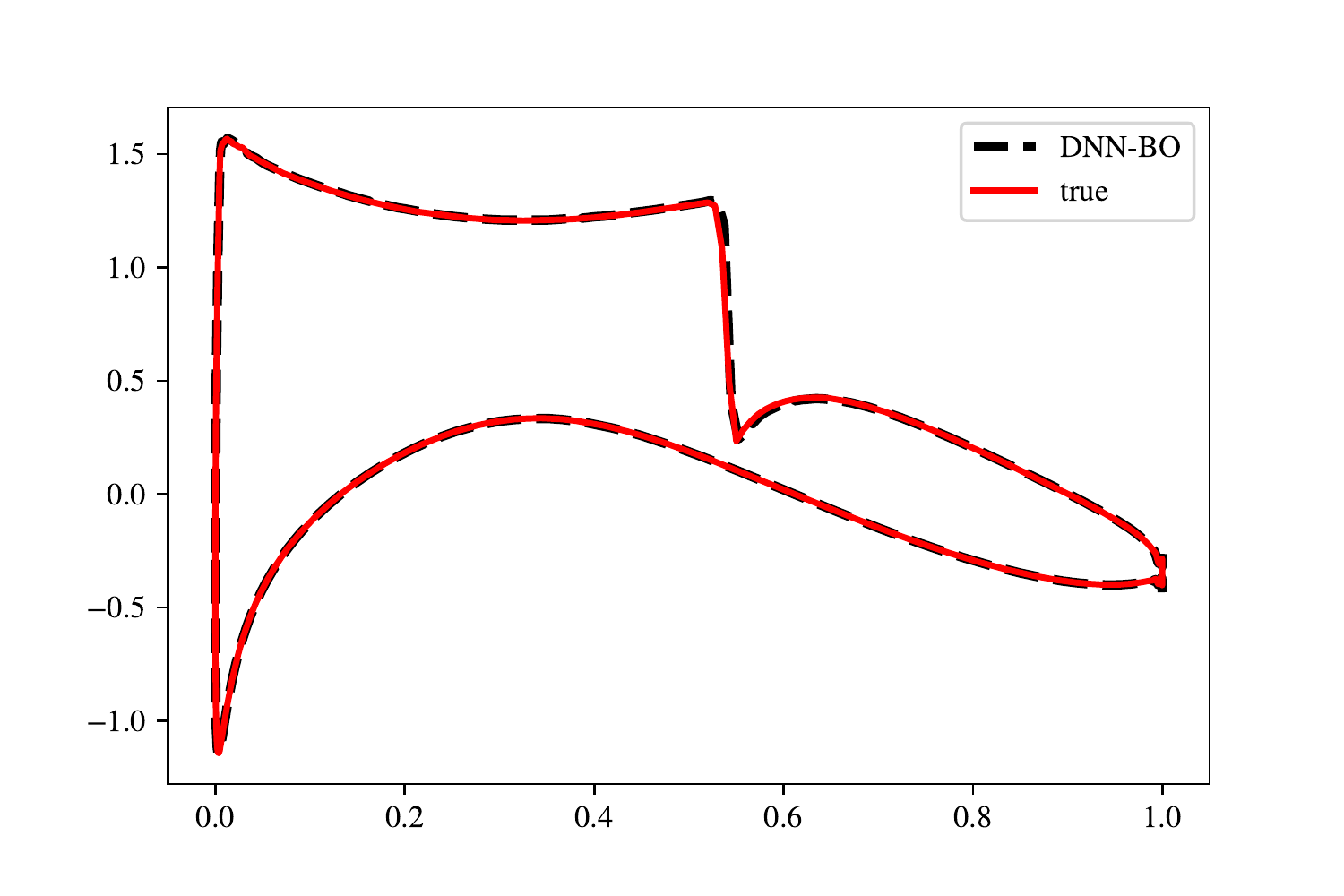}
     \caption{DNN-BO optimized}
     \end{subfigure}\\
    \caption{$C_p$ plots along the airfoil surface for \emph{P1}: unconstrained $C_d$ minimization}
    \label{f:uncon_cpplots}
\end{figure}

The $C_p$ contours for the baseline and optimized airfoil shapes are shown in Figure~\ref{f:uncon_pressure} which helps visualize the aerodynamic effect of the ASO problem. More specifically, the optimization weakens the shock and moves it upstream thereby reducing the drag. Also note that while the DNN overpredicts this effect, the DNN-BO predicts the true globally optimum shape almost exactly. This is further further visualized by the airfoil $C_p$ plots shown in Figure~\ref{f:uncon_cpplots}. Finally, the optimized lift and drag coefficients are summarized in Table~\ref{t:final_cdcl}. Notice that the optimized $C_d$ predicted by DNN is very close to the true optimum however, the $C_l$ is relatively off. This is because the DNN does not encode any dependence structure between the $C_d$ and $C_l$ and hence the predictions are independent of each other. Therefore, as a result of bias in the predictive accuracy, even if the predicted minimum drag is numerically close to the true optimum, the predicted $C_l$ does not necessarily match equally closely with the true $C_l$ at the minimum drag configuration. On the other hand, the $C_d$ and $C_l$ predicted by the DNN-BO almost exactly because the BO framework returns one of the actual observed values as the minimum; see Algorithm~\ref{a:DNN-BO}.

\clearpage
\subsection{\emph{P2}: $C_l$-constrained $C_d$ minimization}
\label{s:con}
\begin{figure}[ht]
    \centering
    \begin{subfigure}{0.5\textwidth}
     \includegraphics[width=\linewidth]{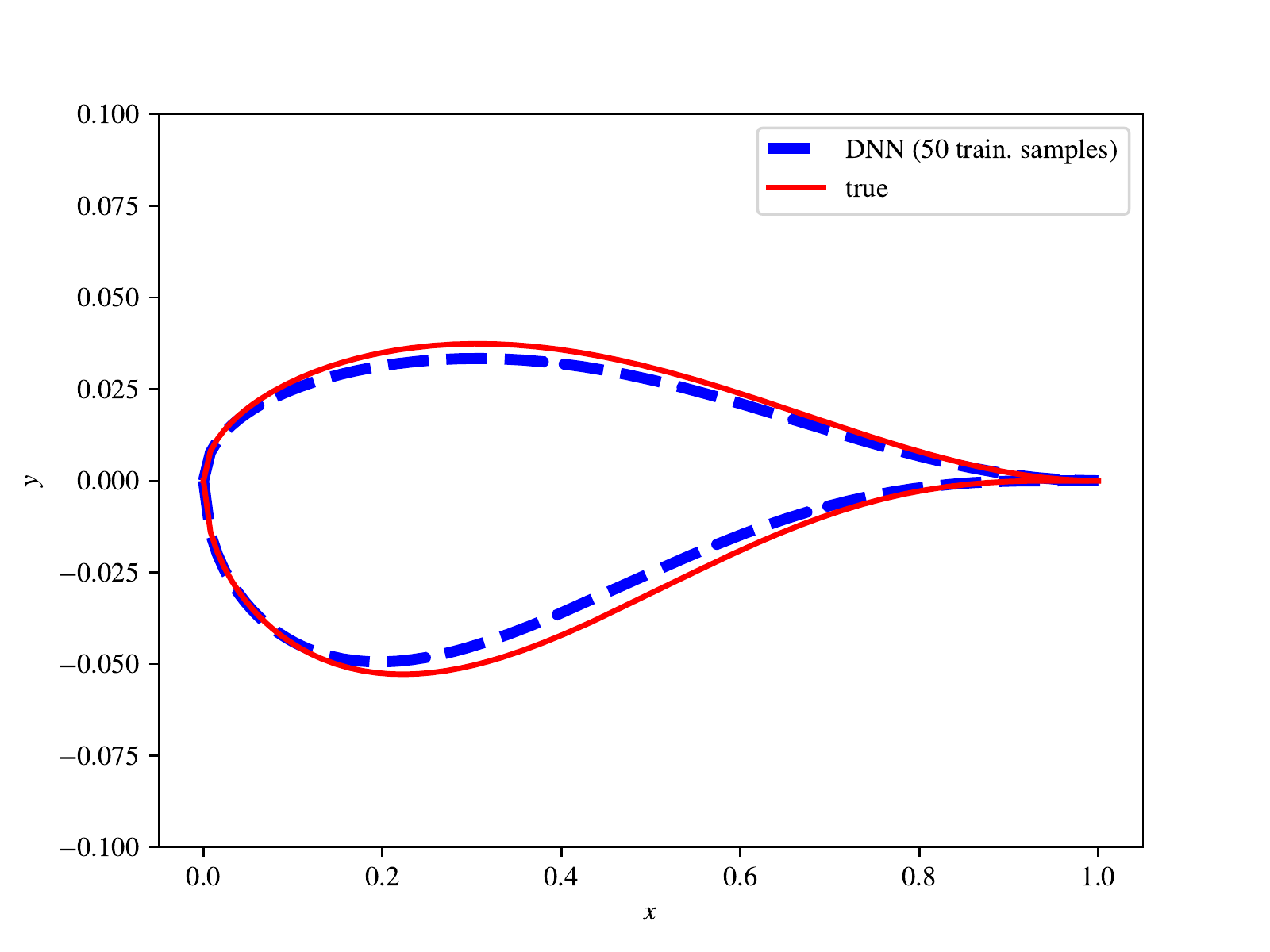}
     \caption{DNN-based optimization (50 training samples)}
     \end{subfigure}%
    \begin{subfigure}{0.5\textwidth}
     \includegraphics[width=\linewidth]{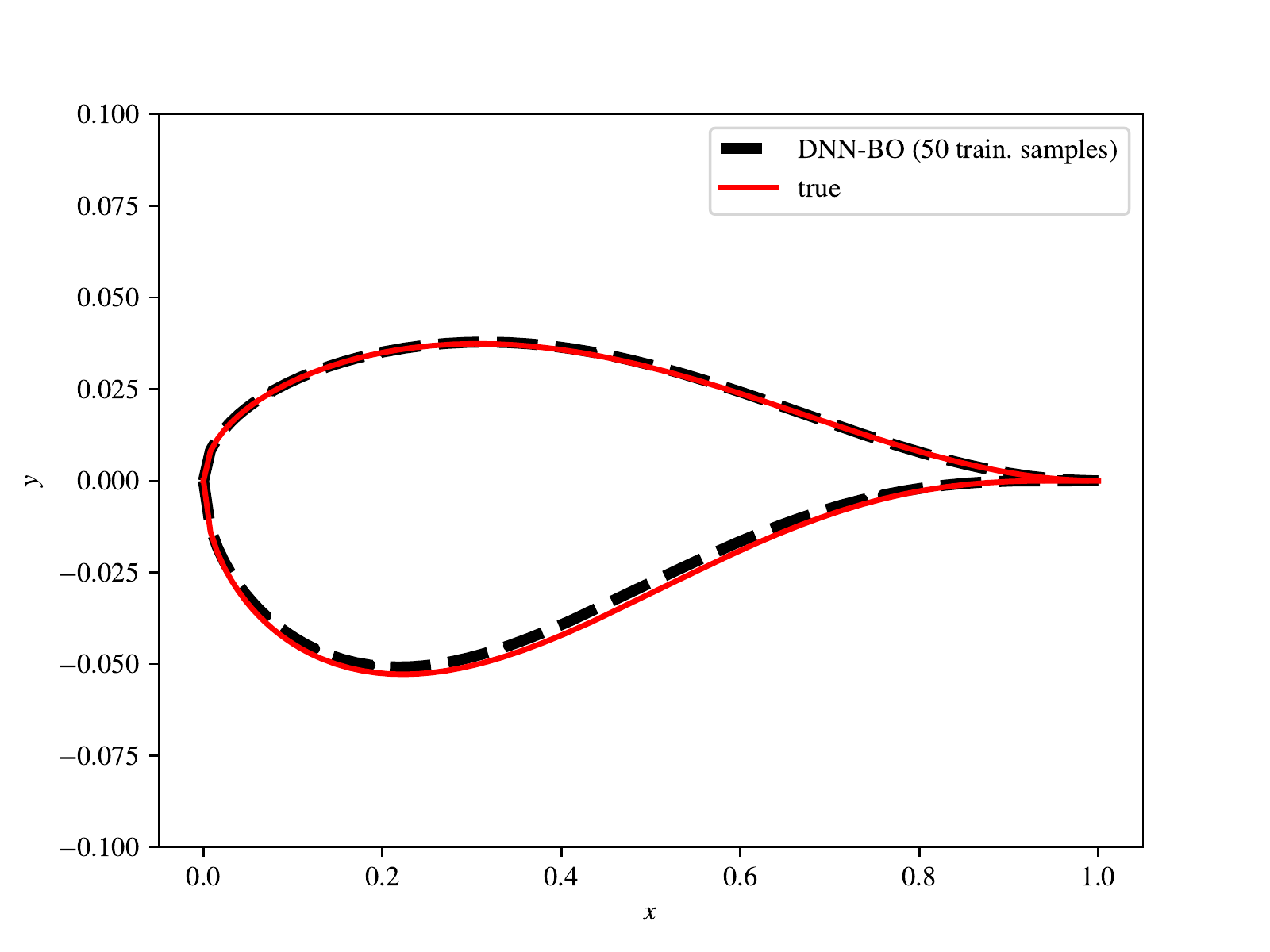}
     \caption{DNN-enhanced Bayesian optimization (50 training samples)}
     \end{subfigure}\\
    \caption{Airfoil shapes for \emph{P2:} $C_l$ constrained $C_d$ minimization}
    \label{f:con}
\end{figure}

The results of the constrained optimization $P2$ are shown in Figure~\ref{f:con}. The constrained optimization is relatively more challenging because independent surrogates are used for the objective function and constraints which could potentially compound the prediction errors and furthermore, an accurate emulation of the feasible region is necessary. Encouraged by the success of the surrogate-based approaches on $P1$, here we make direct comparisons between the DNN and DNN-BO approaches.

The results show that the DNN-based optimization in this case suffers further in accuracy which is partially (but to a good extent) remedied by the DNN-BO. Overall, even though the DNN-BO is unable to replicate the superior predictive accuracy of the unconstrained case, they show a significant improvement over using a trained DNN model.

\begin{figure}[ht]
    \centering
    \begin{subfigure}{0.5\textwidth}
     \includegraphics[width=\linewidth]{Baseline_Copy_of_Scalar_Scene_1.png}
     \caption{Baseline}
     \end{subfigure}%
    \begin{subfigure}{0.5\textwidth}
     \includegraphics[width=\linewidth]{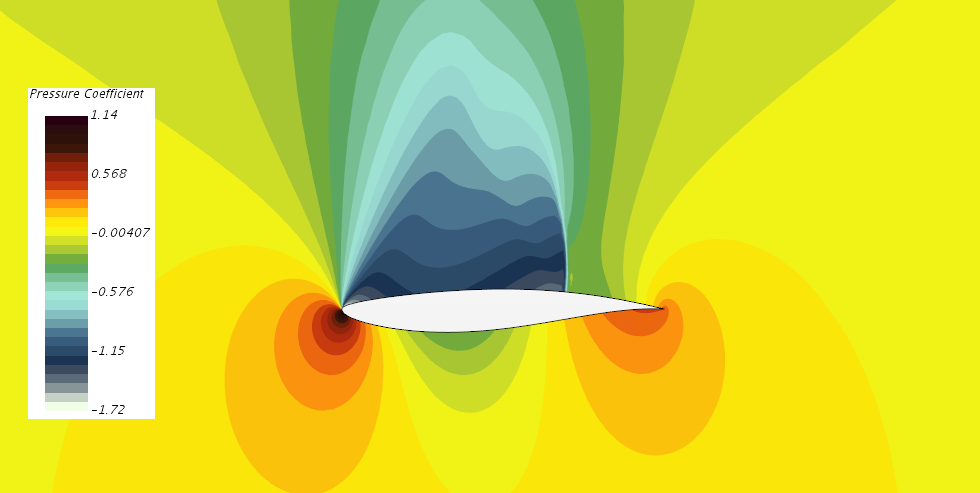}
     \caption{True optimized}
     \end{subfigure}\\
     \begin{subfigure}{0.5\textwidth}
     \includegraphics[width=\linewidth]{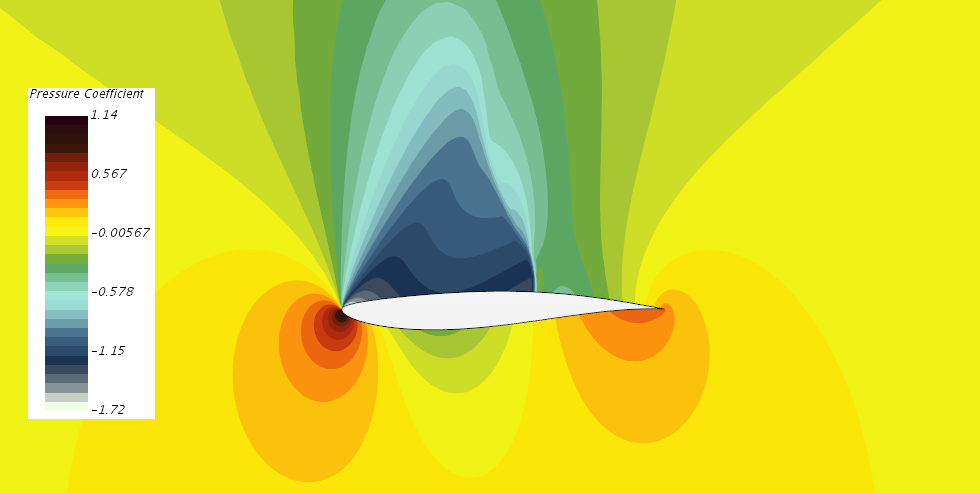}
     \caption{DNN optimized}
     \end{subfigure}%
    \begin{subfigure}{0.5\textwidth}
     \includegraphics[width=\linewidth]{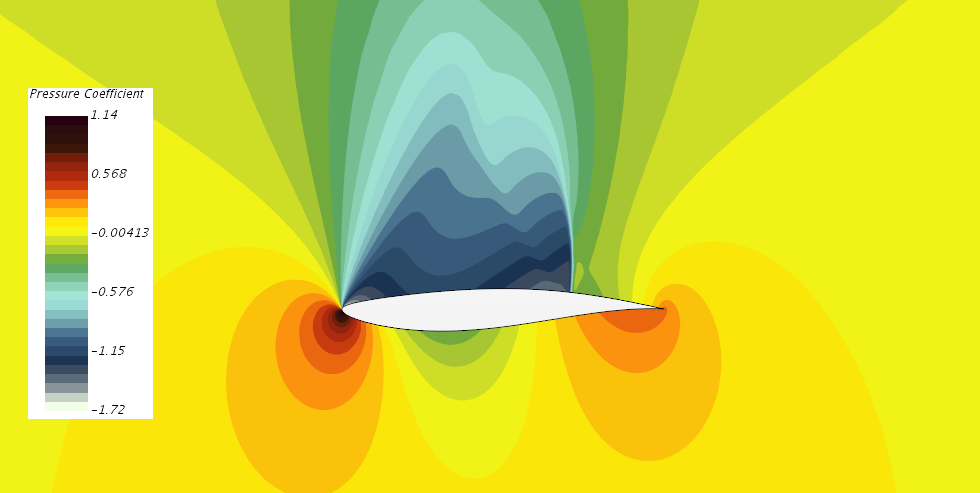}
     \caption{DNN-BO optimized}
     \end{subfigure}\\
    \caption{Pressure distributions for \emph{P2}: $C_l$-constrained $C_d$ minimization}
    \label{f:con_pressure}
\end{figure}

\begin{figure}[ht]
    \centering
         \begin{subfigure}{0.5\textwidth}
     \includegraphics[width=\linewidth]{con_true_base_cp.pdf}
     \caption{baseline}
     \end{subfigure} \\
     \begin{subfigure}{0.5\textwidth}
     \includegraphics[width=\linewidth]{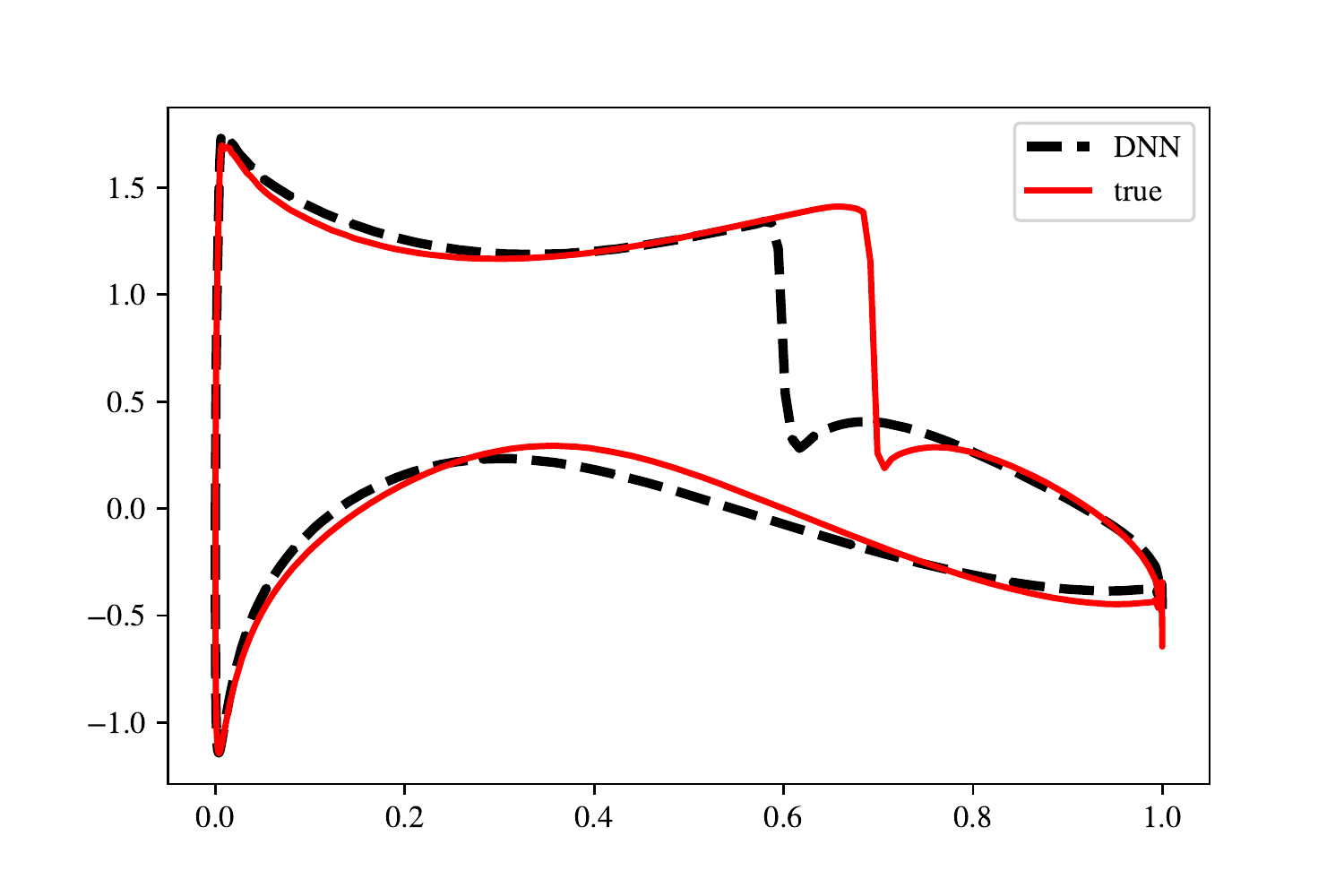}
     \caption{DNN optimized}
     \end{subfigure}%
    \begin{subfigure}{0.5\textwidth}
     \includegraphics[width=\linewidth]{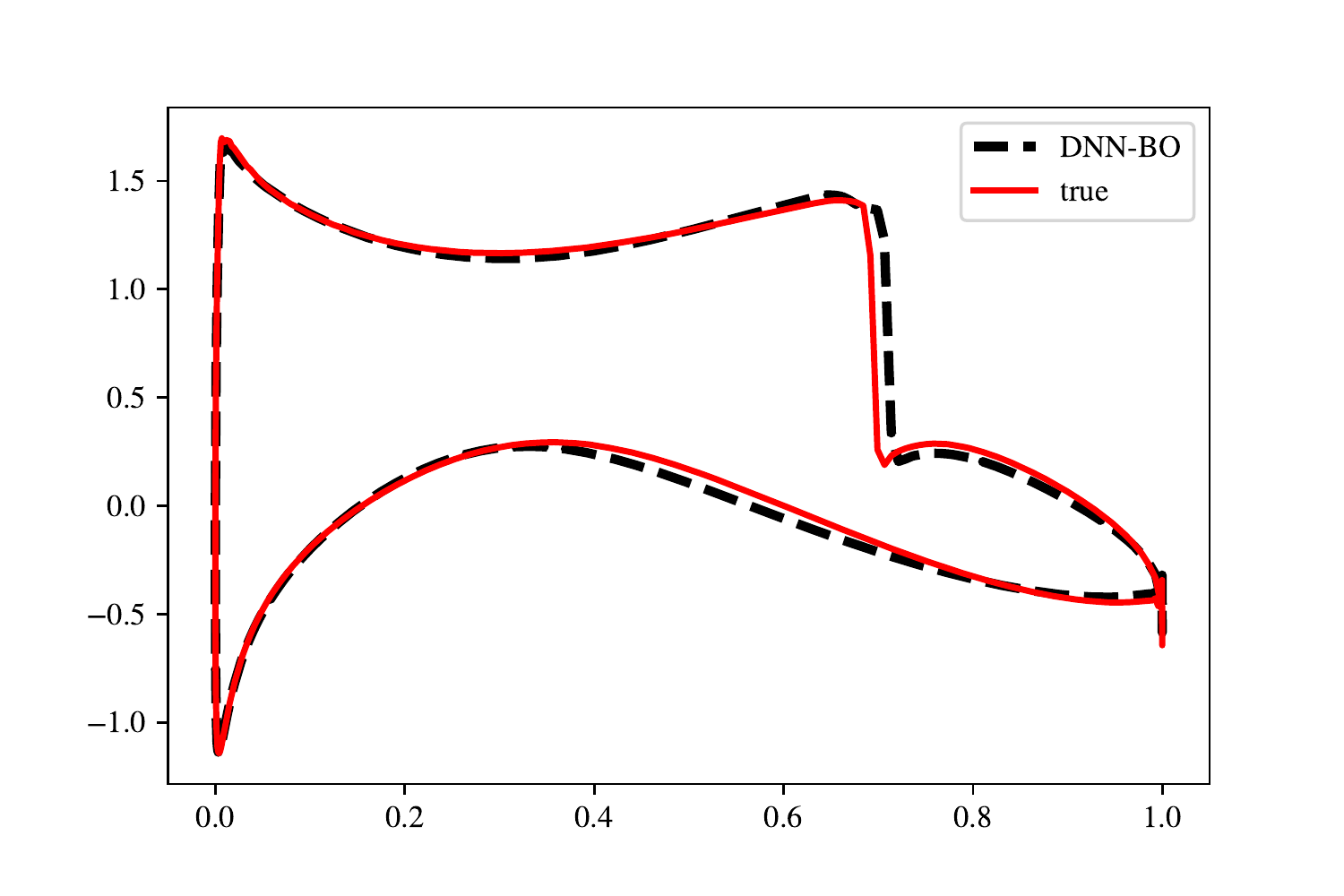}
     \caption{DNN-BO optimized}
     \end{subfigure}\\
    \caption{$C_p$ plots along airfoil surface for \emph{P2}: $C_l$-constrained $C_d$ minimization}
    \label{f:con_cpplots}
\end{figure}

As in the $P1$ results, it is easy to se the accuracy of the DNN-BO approach over the DNN by visualizing the $C_p$ contours, shown in Figure~\ref{f:con_pressure}. It can be seen that the DNN results can lead to  gross overprediction of the shock weakening and upstream displacement compared to the DNN-BO. This is further corroborated via the airfoil $C_p$ plots in Figure~\ref{f:con_cpplots}.

The final lift and drag coefficients for \emph{P2} are shown in Table~\ref{t:final_cdcl}. For the constrained optimization, the final $C_d$ predicted by the DNN is significantly off compared to the true optimum. The primary reason is that in such data-sparse situations, it is a challenge to accurately train a highly parametrized model such as the DNN. This is significantly improved upon by enhancing the DNN with a GP model, as evident from the results of the DNN-BOc. The final $C_l$ for the DNN-BOc is not strictly equal to 1.0 because we use the AL framework where constraint violation is penalized but not entirely avoidable. Due to the rather small training dataset size of 40 points, the trained GP is not accurate enough to predict the true constrained optimum exactly. However, the drastic improvement in accuracy compared to the DNN is quite evident. 

\begin{table}[]
    \centering
    \begin{tabular}{clcc}
    \hline
     & & $C_d$ & $C_l$ \\
     \hline
     \multirow{4}{*}{\emph{P1}: Unconstrained} & DNN & $0.008679$ & $0.87597$ \\
     & DNN-BO & $0.0087$ & $0.8345$ \\
     & Adjoint & $0.0126$ &$0.84666$ \\
     & Truth & $0.0087$ & $0.83454$ \\
     \hline
     \multirow{3}{*}{\emph{P2}: Constrained} & DNN & $0.008679$ & $1.00$ \\
     & DNN-BOc & $0.0144$ & $1.0298$ \\
     & Truth  & $0.0130$ & $1.00$ \\
     \hline
         &  \\
         & 
    \end{tabular}
    \caption{Summary of optimized $C_d$ and $C_l$}
    \label{t:final_cdcl}
\end{table}
\section{Conclusions}
\label{s:conclusion}
We propose machine learning enabled surrogate-based approaches for aerodynamic design optimization with a high-fidelity simulation model particularly, aerodynamic shape optimization. Such problems are traditionally solved via the adjoint method which provides high-quality gradients as a solution to the adjoint equation which can then be used in a gradient-based optimizer. However, adjoint-based approaches can be computationally expensive since they require two expensive, high-fidelity function evaluations for every step in the optimization search. Furthermore, the solution to the adjoint equation needs careful mathematical treatment and can be a challenge to obtain sometimes.

We propose alternatives to the accurate but expensive adjoint method via state of the art machine learning methods. Specifically, we propose construction of surrogate models of the objective function and constraints via a deep neural net (DNN) and a Gaussian process model with DNN prior (DNN-GP). Then gradient-based optimization is performed directly on the DNN and via a Bayesian optimization setting on the DNN-GP (DNN-BO). In both cases, the surrogate model is constructed from a Latin Hypercube design of experiments on the input space that is agnostic to the high-fidelity model. 

Demonstration on the unconstrained drag minimization of the RAE2822 airfoil under inviscid transonic conditions revealed that, the DNN and DNN-BO approaches accurately predict the optimum shape with only $1/12$ as many expensive function evaluations of the high-fidelity model, as the adjoint-based method. Furthermore, we show that we can solve multiple ASO problems with the same surrogate model, by demonstrating it on the $C_l$-constrained $C_d$ minimization. Therefore, without any additional expensive function evaluations, the proposed DNN-BO approach accurately predicts the constrained optimum shape as well. On the other hand, using an adjoint-based approach on multiple optimization problems does not take advantage of the data reuse that our proposed approach does and could have potentially cost an order of magnitude more expensive function evaluations. We demonstrate our approach on two optimization problems but one could further extend the proposed approach to more problems for which the trained machine learning models are valid.

Despite the success of our proposed approach, one possible limitation is that as the input dimension size increases, the required training data size can potentially become intractable, a phoenomenon known as \emph{curse of dimensionality} in machine learning. On the other hand, the adjoint-based method does not suffer from this limitation. A possible remedy for this within our proposed framework is to incorporate a dimensionality reduction e.g., via active subspaces~\cite{constantine2015active, lukaczyk2014active} to identify a more tractable low-dimensional subspace . This is particularly relevant in the ASO problem, where the aerodynamic shape parameters exhibit strong correlations with each other and deformation can be relatively more pronounced in localized regions e.g. the airfoil leading and trailing edges. This is one direction of our future work.

Another direction of our future work is to evaluate a trust-region based optimization framework where the optimization search is performed with a sequence of local surrogate models, which could potentially be constructed with smaller training datasets and yet can provide robust local approximation.

\clearpage
\section*{Acknowledgements}
This material is based upon work supported by the U.S. Department of Energy (DOE), Office of Science, Office of Advanced Scientific Computing Research, under Contract DE-AC02-06CH11357. This research was funded in part and used resources of the Argonne Leadership Computing Facility, which is a DOE Office of Science User Facility supported under Contract DE-AC02-06CH11357. SAR acknowledges the support by Laboratory Directed Research and Development (LDRD) funding from Argonne National Laboratory, provided by the Director, Office of Science, of the U.S. Department of Energy under contract DE-AC02-06CH11357. RM acknowledges support from the Margaret Butler Fellowship at the Argonne Leadership Computing Facility. This paper describes objective technical results and analysis. Any subjective views or opinions that might be expressed in the paper do not necessarily represent the views of the U.S. DOE or the United States Government. Declaration of Interests - None.

\begin{mdframed}
    The submitted manuscript has been created by UChicago Argonne, LLC, Operator of Argonne National Laboratory ("Argonne”). Argonne, a U.S. Department of Energy Office of Science laboratory, is operated under Contract No. DE-AC02-06CH11357. The U.S. Government retains for itself, and others acting on its behalf, a paid-up nonexclusive, irrevocable worldwide license in said article to reproduce, prepare derivative works, distribute copies to the public, and perform publicly and display publicly, by or on behalf of the Government. The Department of Energy will provide public access to these results of federally sponsored research in accordance with the DOE Public Access Plan (http://energy.gov/downloads/doe-public-access-plan).
\end{mdframed}

\bibliographystyle{unsrt}  
\bibliography{References}  

\end{document}